\documentclass[12pt]{article}

\usepackage{amsmath, amssymb, amsthm}
\newcommand {\half} {{1 \over 2}}
\newcommand {\hhalf} {{\textstyle{1 \over 2}}}

\newcommand {\bra}[1] {\left< #1 \right|}
\newcommand {\ket}[1] {\left| #1 \right>}
\newcommand {\expect}[1] {\left< #1 \right>}
\newcommand {\braket}[2] {\left< #1 \mskip1mu\vrule\mskip1mu #2 \right>}

\theoremstyle{remark}

\theoremstyle{definition}

\begin{document}

\title{
        Worldsheet Covariant Path Integral Quantization of 
        Strings
      }
    \author{Andr\'e van Tonder
            \\ \\
            Department of Physics, Brown University \\
            Box 1843, Providence, RI 02912 \\
            andre@het.brown.edu
            }
    \date{September 1, 2006}

    \maketitle

    \begin{abstract}
    
\noindent
We discuss a covariant functional integral approach to the quantization
of the bosonic string.
In contrast to approaches relying on 
non-covariant operator regularizations, interesting operators here 
are true tensor objects with classical transformation laws, even on 
target spaces where the theory has a Weyl anomaly.  
Since no implicit non-covariant gauge choices are involved in 
the definition of the operators, the anomaly
is clearly separated from the issue of operator renormalization and can be
understood in isolation,
instead of infecting the latter as in other approaches.  
Our method is of wider applicability to 
covariant theories that are not Weyl invariant, but where 
covariant tensor operators are desired. 
  
After constructing covariantly regularized vertex
operators, we define a class of background-independent 
path integral measures suitable for string quantization.  We show how gauge invariance of the 
path integral implies the usual physical state conditions
in a very conceptually clean way.  We then 
discuss the construction of the BRST action from first 
principles, obtaining some interesting caveats relating
to its general covariance.  
In our approach, the expected BRST related anomalies are
encoded somewhat differently from other approaches.  
We 
conclude with an unusual but amusing derivation of 
 the value $D= 26$ of the critical dimension.

        %\noindent
        %\rule{4.6in}{1pt}
        %\newline
        %{Keywords:}
        %  Conformal Field Theory,  Path Integral, Anomaly
        %\newline
        %arXiv.org eprint: hep-th/
        %\newline
        %{Brown preprint:} BROWN-HET-1429
        %\newline
        %{PACS:}
        %  11.25.Hf, 11.10.Gh
    \end{abstract}

\section{Introduction}

We discuss the covariant functional integral quantization
\cite{myself} of
the bosonic string.  Our approach is to take the functional 
integral seriously in the spirit of Fujikawa \cite{fujikawa1, fujikawa2},
 and to construct a covariantly regularized
theory based on the methods introduced by Pauli and Villars
\cite{myself, PV, vilenkin, anselmi}. 

Our purpose is twofold.  First, to show how familiar results in 
string theory can be recast in a covariant framework that 
that has conceptual and practical advantages over 
non-covariant approaches to operator regularization.  And second, 
to illustrate in a simple setting 
some techniques for defining background independent functional 
integrals in 
generally covariant theories.  

In our covariant approach, we find the same 
physical anomalies as in existing non-covariant approaches
\cite{eguchi, ooguri, BPZ, difrancesco, polchinski, henkel}.  However, 
the anomalies are manifested in the formalism in a different way
 that is often quite illuminating.  

One of the main practical advantages of the covariant approach over the 
alternative operator approaches is that 
it permits one to construct interesting quantum operators that are 
finite, true covariant tensor objects, even in the presense of
anomalies.
There are no implicit gauge choices incorporated in our regularization, 
as in the non-covariant approaches.  While this distinction is not 
that important for the critical string, it is of interest 
 in the wider
context of quantization of curved space-time theories
that are not necessarily Weyl-invariant.  

The covariant approach directly relates the 
anomalies to a failure of Weyl invariance at the quantum level.
In contrast to the non-covariant approaches, this Weyl
dependence is fully explicit in the action, and is not hidden
in the measure or the regularization. 

We show that anomalies are 
encoded in the fact that the trace of the energy momentum tensor
(which is now a true tensor object) is not
zero, but contributes contact terms when contracted with various
operators. 

In the familiar non-covariant regularizations, it is much more 
difficult to disentangle effects of the quantum gauge 
transformations
from the implicit gauge choices that are made in the 
choice of regularization. In these approaches, operators defined in 
different coordinate frames are typically related not only by
a coordinate transformation but also by a potentially anomalous
gauge transformation.  Disentangling the various contributions
requires some gymnastics, and becomes conceptually quite
involved.  Indeed, it is 
often quite non-obvious whether the coordinate dependency
introduced via a given operator regularization is in fact
equivalent to a choice of gauge.  If it is not, 
any quantum consistency conditions derived from them would be
spurious and meaningless.  
The covariant approach of this paper does not suffer from this 
problem. 

For these reasons, it is our hope that the covariant approach
may be of some conceptual usefulness in the study of conformal
field and string quantization.  Our construction of 
background-independent path integral measures, and of
the BRST action in the covariant approach,
may have 
wider application in generally covariant theories.  

The organization of the paper is as follows:

We first discuss the construction of covariantly regularized vertex
operators.  In contrast to existing approaches, these
operators are true tensors satisfy classical Ward identities with respect
to the energy-momentum tensor, which is a true covariant 
object.  

After discussing the construction of background-independent 
path integral measures, we are ready to move on to string 
quantization.  We show how gauge invariance of the 
path integral implies the usual physical state conditions
in a way that is conceptually quite clean, if somewhat 
technically demanding.  The correspondence with the 
Virasoro conditions is then demonstrated, and is rather
indirect.

We then 
discuss the construction of the BRST action from first 
principles.  Important in the current approach is the issue of background 
invariance of the BRST action, which we discuss rather carefully.  
We also discuss anomalies from the BRST point of view. 
In our approach, the BRST current is a true, non-anomalous
tensor object, and the expected anomaly and physical state conditions are 
encoded differently from the operator approach.  
Their most natural expression in the formalism is in terms of the effective action
and the antibracket.  

For completeness, and because the calculation is sufficiently
different from other approaches to make it interesting, we 
conclude by deriving the value $D= 26$ for the critical dimension.  

This paper relies extensively on the techniques and results of
\cite{myself}. 

\section{Finite vertex operators and states}

In this section we discuss a useful covariant renormalization
of vertex operators in the covariant functional integral approach introduced
in reference \cite{myself}.  We will assume the form of the action 
introduced in that reference, which is the $(X, \chi)$ part of 
the full string action derived later and given by formula (\ref{covBRST}).
Here the $X$ are the matter fields and their Pauli-Villars partners
are denoted by $\chi_i$.

To illustrate how vertex operators may be renormalized in 
a coordinate-invariant way in the Pauli-Villars formalism, 
consider a path-integral calculation of
the state corresponding
to the operator
$$
  e^{ik\left(X + \sum_i\eta_i\chi_i\right)}.
$$ 
Here the $\chi_i$ range over the 
Pauli-Villars regulator fields, and the $\eta_i$ are complex numbers.
Some of the $\chi_i$ are commuting real scalars, and some are 
complex Grassmann scalars \cite{myself}.  We require 
$$
  \eta_i = 0, \quad \textrm{for $\chi_i$ Grassmann}.
$$
We shall show that this expression will provide a full renormalization of the undressed 
insertion
$e^{ikX}$ once 
the masses of the regulator fields and the constants 
$\eta_i$ have been chosen to
make path integrals containing it finite. 

To obtain the state corresponding to this vertex operator, we need
to calculate the path integral 
\begin{align*}
  \int_{(X, \chi)_{\partial D} = (X_b, \chi_b)} 
& [dX]\wedge [d\bar\chi] \wedge [d\chi]\\
   \times\exp\biggl(&-\half\int d^2x\, 
\biggl\{4 \,\partial X\bar\partial X + m^2 X^2 +  \sum_i 
        \left(4\,\partial\bar\chi_i\bar\partial\chi_i 
        +  M_i^2 \chi_i^2\right)\biggr\} \\
      &+ ik\biggl\{X(0) + \sum_i \eta_i\chi_i(0)\biggr\}
  \biggr) 
\end{align*}
where the path integral is taken over configurations of the fields on the
unit disc $D$ with boundary $\partial D$ to obtain a functional of
the boundary configuration $(X_b, \chi_b)$ of the fields.\footnote{Our normalization of the
action corresponds to replacing
   $\alpha' \to {1\over 2\pi}$.}  At this 
point we ignore the ghost contribution, which will be discussed later.

In \cite{myself}, a careful definition of the path integral measure was 
provided.  
It was shown there that the full measure, after including the Pauli-Villars
contributions, is invariant  under arbitrary variations of the metric, 
and also separately under arbitrary diffeomorphisms acting only on the 
fields.  Putting 
these two facts together, it follows that the measure is 
invariant under full diffeomorphisms and Weyl transformations, 
a desirable property for a conformal field theory.  
Note that this measure differs from the
usual Fujikawa measure for $X$ only, which is not conformally invariant.

Writing $X = X_{c} + \bar X$ where $X_c$ is the extremum of the term in the 
exponent
with $X_c(\partial D) = X_b$, and $\bar X(\partial D) = 0$, the path 
integral over $\bar X$ is Gaussian and contributes an overall normalization,
and we obtain the matter contribution
\begin{align}
 Z\, e^{- S(X_c) + ik\, X_c(0)}  \label{PIresult}
\end{align}
The stationary configuration $X_c$ is obtained by varying $X$ in the bulk of $D$ to 
get
$$
  -4 \,\partial\bar\partial X_c + m^2 X_c - {ik}\, \delta^2(x) = 0.
$$
This is solved by 
$$
  X_c = X_0 + \sum_{n>0}\left(z^nX_n + \bar z^n X_{-n}\right)
           - {ik\over 4\pi}\, \ln m^2 z\bar z
$$
in the limit as $m \to 0$, remembering that $K_0(mr) \to - \ln mr$ as 
$m \to 0$, or simply noting that $\partial\bar\partial \ln z\bar z = \pi \delta^2 (x)$.  

Note that the extremum $X_c$ is complex, so that the above decomposition 
$X = X_c + \bar X$, with 
$\bar X$ real, describes
an integration path in field space that has been displaced away from 
the real line.  It is straightforward to convince oneself that,
as in the  one-dimensional analogue
$$
  \int_{-\infty}^\infty dz\, e^{-az^2} = 
 \int_{-\infty+ ib}^{\infty+ ib} dz\, e^{-az^2},
$$
such a displacement 
does not change the value of the integral, 
since no poles are crossed and the 
Gaussian decays sufficiently rapidly.

Note, however, that the boundary values
$X_c (|z| = 1) = X_b (z)$ of the field are constrained to be real,
which implies that
$$X_0 = {ik\over 4\pi} \ln m^2 + x_0,$$
where $x_0$ is real, and we may write
$$
  X_c = x_0 + \sum_{n>0}\left(z^nX_n + \bar z^n X_{-n}\right)
           - {ik\over 4\pi}\, \ln z\bar z
$$
Inserting the stationary solution, the path integral becomes 
\begin{align*}
  Z \, e^{ik x_0}  \exp\left\{-2\pi \sum_{n>0} n X_n X_{-n}\right\}\,
       \exp \left\{-\half\,{(ik)^2\over 4\pi}\, \ln 0^2\right\}.
\end{align*}
The first two exponentials are just proportional to \cite{polchinski}
$$  
  e^{ikx_0} \ket{0}_X.  
$$
The term indicated by $\ln 0^2$ in the last exponential would be 
divergent in the absence of the Pauli-Villars contributions.  
However, it is neatly canceled once we include the latter.  
For the Pauli-Villars fields, the stationary solutions
may be approximated for large $M_i$ by Bessel functions as
$$
   \chi_i^c \to \chi^i_0 + \sum_{n>0}\left(z^n\chi^i_n + \bar z^n \chi^i_{-n}\right)
     + {ik\over 4\pi}\, 2\eta_i\,K_0 (M_i r).
$$
In the infinite mass limit, the Bessel functions tend 
to
zero on the boundary.  As a result, we get no 
$\exp {ik\chi^i_0}$ contributions, in contrast with the
field $X$ above.  The path integral becomes  
\begin{align*}
  Z' \, \prod_i\exp\left\{-2\pi \sum_{n>0} n \chi^i_n \chi^i_{-n}\right\}\,
       \exp \left\{-\half\,{(ik)^2\over 4\pi}\, 
   \biggl(-\sum_i 2  \, \eta_i^2\, K_0(M_i 0)
   \biggr)\right\}.
\end{align*} 
The first exponent is proportional to the vacuum
$\ket{0}_{PV}$ of the Pauli-Villars fields. 
The potentially divergent terms in the exponent contribute 
\begin{align*}
  \lim_{r\to 0}
  \biggl(
     \ln r^2 &-\sum_i 2  \, \eta_i^2\, K_0(M_i r)
   \biggr) \\
   &= \lim_{r\to 0}
\left(\ln r^2 +\sum_i  \eta_i^2\, \ln M_i^2 r^2
\right) \\
  &=
   \sum_i \eta_i^2\, \ln M_i^2
   +\lim_{r\to 0} \left(1 +\sum_i  \eta_i^2\right) \ln r^2,
\end{align*}
remembering that $K_0 (Mr) \to - \ln Mr$ as $r\to 0$.
Although we have arbitrarily introduced an short-distance
cutoff $r$, the Pauli-Villars result is independent of the
precise cutoff method used once we take the continuum limit.  
It is indeed this property of the Pauli-Villars regularization
that makes it suitable as a non-perturbative, 
coordinate-invariant regularization.  

We now impose  the conditions
\begin{align}
   0 &= 1 + \sum_i \eta_i^2, \label{PV1}\\
   0 &= \sum_i \eta_i^2\ln {M_i^2\over \mu^2}, \label{PV2}
\end{align}
on the coefficients $\eta_i$ and the Pauli-Villars masses $M_i$.  
Here $\mu$ is an arbitrary \textit{finite} renormalization scale.  
These conditions can always be satisfied while
taking $M_i\to\infty$ as long as there are enough 
Pauli-Villars fields. 
Since the parameter $\mu$ constrains the way we take the $M_i\to \infty$ 
limit, the resulting path integral measure depends implicitly 
on $\mu$.  In fact, we are really defining a one-parameter
family of path integral measures depending on $\mu$.  

The above expression is then finite
\begin{align*}
 \sum_i \eta_i^2\, \ln M_i^2 =  -\ln \mu^2,
\end{align*}
and the
final result of the path integral is proportional to
\begin{align*}
       \mu^{-k^2/4\pi} \, e^{ikx_0}\ket{0}_X \otimes\ket{0}_{PV}
     \equiv \ket {k}
\end{align*}
Although the $\mu$-dependent prefactor is perfectly finite, 
physical states should not depend on the renormalization scale.   
We can indeed compensate the prefactor by instead 
considering the finitely renormalized insertion
\begin{align}
 \mu^{k^2/4\pi}
  e^{ik\left(X + \sum_i\eta_i\chi_i\right)}
\label{renormalized}
\end{align}
whose correlation functions will be independent of $\mu$.

\section{Two-point functions}

Let us calculate the two-point function
$$
  \expect{e^{ik_1 \left(X(w) + \sum_i \eta_i\chi_i(w)\right)}\, 
      e^{ik_2 \left(X(0) + \sum_i \eta_i\chi_i(0)\right)}}
$$
via a path integral.  The presence of the Pauli-Villars terms in 
the exponents will make the result finite without any additional
renormalization.  

We proceed as in the previous section.  The stationary solutions 
now have two sources and the boundary conditions are different 
from those in the previous section.  On 
the plane with vanishing boundary conditions on $X$ and $\chi_i$ at infinity, the 
solutions are
$$
  X_c =  - {ik\over 4\pi}\, \ln m^2 z\bar z
         - {ik\over 4\pi}\, \ln m^2 (z - w)\overline {(z - w)}
$$
and 
$$
   \chi_i^c = {ik\over 4\pi}\, 2\eta_i\,K_0 (M_i |z|) 
             + {ik\over 4\pi}\, 2\eta_i\,K_0 (M_i |z-w|). 
$$
Similar to (\ref{PIresult}), the result of the path integration is
$$
 Z\, e^{- S(X_c) + ik_1\, X_c(w) + ik_2 \, X_c (0)}
$$
for the matter contribution, and similarly for the Pauli-Villars fields.
After inserting the stationary solutions into this expression, we obtain
\begin{align*}
  Z  \exp \biggl\{- \half\,{1\over 4\pi} 
   \bigg[(ik_1)^2 \lim_{r\to 0}&\biggl(\ln m^2 r^2  -\sum_i 2  \, \bar \eta_i^2\, K_0(M_i r)\biggr)
\\
+ 2 \,(ik_1)(ik_2) &\biggl(\ln m^2 w\bar w -\sum_i 2  \, \bar \eta_i^2\, K_0(M_i |w|)\biggr) \\
 + {(ik_2)^2} \lim_{r\to 0}&\biggl(\ln m^2 r^2  -\sum_i 2  \, \bar \eta_i^2\, K_0(M_i r)\biggr)\biggr]\biggr\}.
\end{align*}
Assuming $w\ne 0$, we now use
$$
  \lim_{M_i \to \infty}  K_0(M_i |w|) = 0.
$$
The total coefficient of $\ln r$ is 
$$
  1 + \sum_i \eta_i^2  = 0,
$$
by the Pauli-Villars condition (\ref{PV1}).
We obtain
\begin{align*}
  &Z  \exp \left\{ \half\,{1\over 4\pi}
    \left(
                 (k_1 + k_2)^2\ln m^2 + 2\, k_1 k_2\ln \bar w w 
              + (k_1^2 + k_2^2) \sum_i \eta_i^2 \ln M_i^2 
   \right)
          \right\} \\
&\qquad = Z\, m^{ (k_1 + k_2)^2/4\pi}\, 
    \mu^{-k_1^2/ 4\pi}\mu^{-k_2^2/ 4\pi}\, |w|^{k_1 k_2/ 2\pi},
\end{align*}
where we have used the Pauli-Villars condition (\ref{PV2}).

We see that, as $m\to 0$, the correlation function vanishes unless
$
  k_1 + k_2 = 0
$.
The final result is finite, and is given by
\begin{align*}
&\expect{e^{ik_1 \left(X(w) + \sum_i \eta_i\chi_i(w)\right)}\, 
      e^{ik_2 \left(X(0) + \sum_i \eta_i\chi_i(0)\right)}} \\
   &\qquad  = \expect{1}
    \left\{
      \begin{array}{ll}
         0 &  \textrm{if $0 \ne k_1+ k_2$}\\
         \mu^{-k_1^2/4\pi} \, \mu^{-k_2^2/4\pi}\,|w|^{k_1 k_2/ 2\pi}
          & \textrm{if $0 = k_1+ k_2$}
      \end{array}
    \right.
\end{align*}
As discussed in the previous section, the $\mu$-dependent prefactors 
can be trivially compensated by a finite renormalization of the vertices, 
in which case the result becomes independent of $\mu$.  

So far we have worked on the plane.  
We may add the point at infinity, in which case 
the only modification to the above analysis is the existence of a 
zero mode $\bar X_0$ over which we should integrate in the 
path integral.  This would contribute an additional factor 
$$
 \int d\bar X_0\, e^{ik_1 \bar X_0 + ik_2 \bar X_0} = 2\pi\, \delta(k_1 + k_2),
$$
to the above result.

\section{General vertex operators}

We now consider the construction of higher vertex operators
in the covariant approach.  These are obtained by 
multiplying derivatives of the fields with the tachyon.  For 
example, the matter part of the graviton is of the form
$$
  \partial X\, \bar \partial X \,e^{ikX}, 
$$
where we have suppressed target space indices.  
In the path integral, these bare insertions are not finite 
due to self-contractions.  In the  
covariant Pauli-Villars approach, we obtain a 
finite insertion by considering instead
$$
  \partial \left(X + \sum_i \eta_i \chi_i\right)\, 
\bar\partial \left(X + \sum_i \eta_i \chi_i\right)\,
\,e^{ik\left(X + \sum_i \eta_i \chi_i\right)}, 
$$ 
In addition to self-contractions already present in the 
exponential, which were shown to be finite in the previous sections, 
this insertion has derivatives of self-contractions of the form
\begin{align}
  &\expect{\partial\left(X + \sum_i \eta_i \chi_i\right)\,
          \left(X + \sum_i \eta_i \chi_i\right)}, \nonumber\\ 
&\expect{\bar \partial\left(X + \sum_i \eta_i \chi_i\right)\,
          \left(X + \sum_i \eta_i \chi_i\right)}, \nonumber\\ 
&\expect{\partial\left(X + \sum_i \eta_i \chi_i\right)\,
          \bar\partial \left(X + \sum_i \eta_i \chi_i\right)}.  \label {deriv}
\end{align}
But these are finite.  Indeed, we have for small $z$,
\begin{align*}
&\expect{ \left(X + \sum_i \eta_i \chi_i\right)_z\,
          \left(X + \sum_i \eta_i \chi_i\right)_0} \\
&\quad\to -{1\over 4\pi}
     \left(\ln m^2 \bar z z + \sum_i \eta_i^2 \ln M_i^2 \bar z z\right) \\
&\quad= \left(\ln m^2 + \sum_i \eta_i^2 \ln M_i^2\right)
    + \left(1 + \sum_i \eta_i^2 \right) \ln \bar z z \\
&\quad= \ln {m^2\over \mu^2} + 0,
\end{align*}
becoming independent of $z$ as $z\to 0$
by the two previously introduced conditions on $\eta_i$ and $M_i$.  
This diverges as $m\to 0$, but taking a derivative, we see that the 
desired contractions  
(\ref{deriv}) are finite, indeed zero,
 independent of $m$, and the result follows. 

This is easily generalized, so that the replacement 
$$
  X \to X + \sum_i \eta_i \chi_i
$$
is sufficient to render the entire tower of vertex insertions
finite.

\section{Covariant Ward identities}
\label{wardsection}

In the previous sections, we saw that the full 
vertex insertion, 
including the Pauli-Villars contributions,
 is finite under suitable conditions on the Pauli-Villars 
masses.  We will now show that the tachyon vertex satisfies 
the classical \textit{scalar} Ward identities with respect to 
coordinate transformations.  We will then relate these to the 
more familiar anomalous Ward identities found in the 
non-covariant operator 
formalism.    

The change of variables formula for the path integral under
a deformation  $X \to X^\lambda = X \circ f_{-\lambda}$ 
of the dynamic fields and their
Pauli-Villars partners along a vector field $v$ generating the
flow $x \to f_\lambda(x)$,
where $\lambda$ is a real parameter along the flow, can 
be written as \cite{myself}
\begin{equation}
  \int [dX]^\lambda_\textit{PV} \,V^\lambda(x)\, e^{-S(X^\lambda, \bar \chi^\lambda, \chi^\lambda)}
    = \int [dX]_\mathit{PV} \,V(x)\, e^{-S(X, \bar \chi, \chi)},
\label{change}
\end{equation}
where 
\begin{align*}
  [dX]_{PV} &\equiv [dX]\wedge[d\bar\chi]\wedge [d\chi], 
\end{align*}
 and 
$V^\lambda \equiv V(X^\lambda, \bar\chi^\lambda, \chi^\lambda)$.  
It should be emphasized that we are not deforming the metric, so that
the above transformation would be a classical symmetry only if $v$ is
conformal.  However, the Ward identities below will be valid for arbitrary
deformations $v$ satisfying suitable boundary conditions.  
For example, on the plane, derivatives of $v$ should vanish sufficiently fast at infinity,
which excludes globally conformal $v$.

The Ward identities are obtained from this formula by 
differentiating with respect to $\lambda$.  
As shown in \cite{myself}, the full measure $[dX]^\lambda_\mathit{PV}$, 
due to the inclusion of the
Pauli-Villars fields, is invariant under the deformation 
$(X, \bar\chi, \chi) \to (X^\lambda, \bar\chi^\lambda, \chi^\lambda)$, and so independent
of $\lambda$.  We therefore obtain, after differentiation, 
\begin{align}
   \expect{{d\over d\lambda}\, V^\lambda(x) \cdots}
     + {1\over 4\pi} \int d^2y \, \sqrt{g}\,h^{kl}(y)\, \expect{T^\lambda_{kl}(y)\,
         V^\lambda(x) \cdots}  =0, \label{ddlambdaV}
\end{align}
where $T^{kl}$ is the full energy-momentum tensor including the Pauli-Villars
contribution, 
\begin{align}
  {d\over d\lambda}\, V^\lambda(x) &= -\mathcal{L}_v  V^\lambda (x)
      + {\delta V^\lambda\over g^{ij}}\, \mathcal{L}_v g^{ij} (x) \nonumber \\
   &= -\mathcal{L}_v  V^\lambda (x)
        + {\delta V^\lambda\over g^{ij}}\, h^{ij} (x)
\label{lvV}
\end{align}
where $\mathcal{L}_v$ denotes the Lie derivative,
and 
\begin{align*}
  h^{kl} \equiv -\nabla^k v^l - \nabla^l v^k.
\end{align*}
Since the 
metric is not being varied under $v$,
operators depending on the metric need the second term on the 
right hand side of (\ref{lvV}) to compensate contributions proportional
to $-\mathcal{L}_v g_{ij}$. 

It is important to note that the transformation $-\mathcal L_v V$ for 
$V$ appearing
in the Ward identity is the \textit{classical} one, 
despite the fact
that $V$ is a fully renormalized, finite insertion.  This is
not what one might naively expect from previous acquaintance with the operator 
formalism, where the Ward identity for $V$ has an anomalous term,
corresponding to the anomalous dimension of the operator.  

However, further thought shows that there 
 is no contradiction.  This formula is in fact correct 
in the full quantum theory, and
is consistent with the operator formalism.  To understand this,
let us specialize to the plane, where $dw\,d\bar w\sqrt g = d^2w$, 
and consider the example
$$
V \equiv e^{ik\left(X+\sum_i \eta_i\chi_i\right)},
$$ 
for which the $\delta V \over \delta g^{ij}$ term vanishes.   
The above then becomes 
\begin{align}
   \expect{(v^z\partial + v^{\bar z}\bar\partial)\, V_z \cdots} 
  = 
    & - {1\over \pi} \int d^2w \,\bar\partial v^w\,\expect{T_{ww}\,
               V_z \cdots} \nonumber \\
  & 
     - {1\over \pi} \int d^2w \,(\partial v^w + \bar\partial v^{\bar w})\,\expect{T_{w\bar w}\,
               V_z \cdots} \nonumber \\
  &  - {1\over \pi} \int d^2w\, \partial v^{\bar w}\,\expect{T_{\bar w\bar w}\,
               V_z\cdots}. \label {vward}
\end{align}
We shall assume growth conditions on $v$ so that the integrals on the right
hand side exist and may be partially integrated without surface 
contributions.  For example, on the plane, $v$ may at most 
tend to a constant vector field at infinity.  
When we add the point at infinity to obtain the sphere, $v$ must be 
everywhere defined, and includes the fields
 $v^z = a + bz + c z^2$ generating the group $PSL(2, \mathbf{C})$
of unimodular M\"obius transformations.  However, we emphasize that 
this formula is valid for general $v$, not just holomorphic or conformal 
$v$.

The consistency of the above result with the usual operator formalism may be 
understood by noticing that, although classically
$T_{w\bar w} = 0$, in a covariantly regularized quantum field theory
$T_{w\bar w}$ can lead to contact terms in 
expectation values \cite{myself, gawedzki, cappelli1, cappelli2, forte}.  
Such contact terms were obtained from axiomatic considerations
in \cite{gawedzki, cappelli1, cappelli2} and were discussed and 
calculated in much detail in \cite{myself} using a covariant Pauli-Villars
regularization.   A similar calculation
from first principles will be done below, but before we do that, we 
illustrate a simpler, though indirect, derivation of the contact terms.  

This proceeds by noting that the Pauli-Villars-regularized objects
are by construction coordinate-invariant and finite.  The above derivation 
of the Ward identity is therefore rigorous, and may be used as 
a starting point for inferring the contact terms. 
Taylor-expanding the exponential and 
performing single and double contractions with $T_{zz}$, we obtain, 
in the limit of infinite Pauli-Villars mass,
\cite{polchinski}
\begin{align*}
  \expect{T_{ww}\, e^{ik\left(X+ \sum \eta_i\chi_i\right)(z)}\cdots} &= {{\alpha' k^2/4}\over (w - z)^2}
     \expect{ e^{ik\left(X + \sum \eta_i\chi_i\right)(z)}\cdots} \\
  &\quad
                   + {1\over w-z} \,\partial_z\expect{e^{ik\left(X+ \sum \eta_i\chi_i\right)(z)}\cdots}
   +\cdots.
\end{align*} 
and similarly for the product with $T_{\bar w \bar w}$.  The final
ellipsis stands for any terms that do not arise from self-contractions
among fields in $T_{ww}\, e^{ikX_z}$, and may also contain terms 
proportional to $\chi_i(z)$ whose matrix elements will vanish, 
in the limit of infinite Pauli-Villars mass, as long as no additional 
insertions are at $z$.  For their explicit form, see the calculation 
later in this section.  As above, $T_{ww}$ denotes the full 
energy-momentum tensor including the contributions
 of the Pauli-Villars auxiliary fields.

The above calculation is familiar from the operator 
formalism, but deserves a few comments in the present context.  First,
note that the insertions are already regularized, since
divergences due to contractions of fields at the same point are cancelled by 
the contributions of the Pauli-Villars fields as in the previous section and 
reference \cite{myself}.  It is important, though,
to make sure that we are not overlooking contact terms due to contractions
of Pauli-Villars terms in $T_{ww}$ with $e^{ikX_z}$.  
This is easily verified for single contractions, where the relevant terms 
would be  
proportional to $\expect {\partial_w \chi\, \bar\chi} \sim \partial_w K_0(Mr)$.  Since the Bessel function $K_0(Mr)$
is positive and has area $2\pi/M^2$ in two dimensions, it tends to the 
zero distribution as we take $M \to \infty$, and therefore so does its 
derivative $\partial_w K_0(Mr)$.  Slightly less obvious are the 
contributions due to double contractions.  A typical term
is proportional to 
$\expect{(\partial_w\chi)^2 \, \chi^2}$, whose Fourier transform can 
be calculated as 
in \cite{myself} to be proportional to 
$$
  {1\over M^2}\int_{2M}^\infty d\mu \,c(\mu, M)\,{\mu^2\, (p_1 - ip_2)^2
     \over p^2 + \mu^2},
$$
where $c(\mu, M)$ is a (spectral) function of unit area with support on 
$[2M, \infty)$.  Due to the lower
bound on the integration and the unit area property, this indeed becomes
$(p_1 - ip_2)^2/M^2 \to 0$ in the limit as $M\to \infty$.

We may now obtain the contact terms by inserting the above result 
into the Ward identity (\ref{vward}), performing partial 
integrations,\footnote{The surface terms 
arising from partial integrations will be 
zero if $v$ goes to a constant at infinity.  Also note that as $w \to z$, the 
distributions $1/(w-z)^n$ are integrable and do not spoil our ability 
to perform partial integrations.}
 and using 
$$
  \pi \,\delta^2 (w - z) = \partial_w \partial_{\bar w}\, \ln |w - z|^2,
$$
we find 
$$
   {1\over \pi}\int d^2w\, (\partial v^w + \bar\partial v^{\bar w})\expect{T_{w\bar w}\,
               V_z \cdots} 
   = {\alpha' k^2\over4}\,  (\partial v^z + \bar\partial v^{\bar z})
   \expect{V_z\cdots } + \cdots.
$$
Since this is true for general $v$, we obtain
\begin{align}
  T_{w\bar w}\,
               V_z   = {\pi}\,{\alpha' k^2\over 4} \,\delta^2(w - z) 
   \,V_z + \cdots.  \label{contact} 
\end{align}
Since this formula is so important in what follows, and since it 
is not usually calculated in more familiar regularization schemes, 
we now verify it by a direct calculation.  We have 
\begin{align}
  &T_{w\bar w}\, e^{ik\left(X(z) + \sum_i \eta_i \chi_i\right)} \nonumber\\
   &\qquad = {\pi\over 2} \left(m^2\phi^2 + \sum_i M_i^2 \chi_i^2\right)\,
     e^{ik\left(X(z) + \sum_i \eta_i \chi_i(z)\right)} \nonumber\\
  &\qquad = {\pi\over 2} \biggl\{{(ik)^2\over 2!}
     \biggl( m^2\expect{X^2(w)\,X^2(z)} 
           + \sum_i \eta_i^2 M_i^2 \expect{\chi_i^2(w) \, \chi_i^2(z)}\biggr)
\nonumber \\
&\qquad \qquad  + ik\biggl(m^2 \, X(w) \expect{X(w)\,X(z)}
               + \sum_i \eta_i M_i^2\, \chi_i(w) \expect{\chi_i(w)\,\chi_i(z)}\biggr) \biggr\} \times\nonumber\\
&\quad \qquad\times e^{ik\left(X(z) + \sum_i \eta_i \chi_i(z)\right)} 
 + \cdots,
\label {fullcontact}
\end{align}
where the ellipsis denotes the non-contracted remainder.  
The double contractions in the first line were already partially calculated
in \cite{myself}.  The result, writing $w \equiv x_1 + ix_2$ and 
$z \equiv y_1 + iy_2$, is
\begin{align}
  &{1\over (\pi/2)}\, {(ik)^2\over 2!}
   \int {d^2 p\over (2\pi)^2}\, e^{-ip\cdot (x - y)}\,
     {1\over 16}\cdot {\pi\over 3}\left( \int_{2m}^\infty d\mu\,
       {c(\mu, m)\over m^2}\,{\mu^4 \over {p^2 + \mu^2}} + \mathit{PV}\right),
\label{double}
\end{align}
where the unit area spectral function is explicitly given by
$$
   c(\mu, m) \equiv  {24\, m^4\over \mu^5\,\sqrt{1 - {4m^2/ \mu^2}}}\,
         \theta (\mu - 2m).
$$ 
Writing
\begin{align}
 {\mu^4 \over {p^2 + \mu^2}} = \mu^2 -  {\mu^2 p^2 \over {p^2 + \mu^2}},
\label{identity}
\end{align}
we obtain a term
$$
  \int_{2m}^\infty d\mu\,{c(\mu, m)\over m^2}\, \mu^2 
   + \sum_i \eta_i^2\, \int_{2M_i}^\infty d\mu\,{c(\mu, M_i)\over M_i^2}\, \mu^2. 
$$
Changing variables from $\mu$ to $\nu \equiv 2m\mu$ and $\nu \equiv 2M_i\mu$ 
respectively removes all mass-dependence from this formula, and the result
is proportional to
$$
  1 + \sum_i \eta_i^2 = 0.
$$
The second term in the identity (\ref{identity}) contributes 
\begin{align*}
  &\int_{2m}^\infty d\mu\,{c(\mu, m)\over m^2}\, 
     {\mu^2 p^2 \over {p^2 + \mu^2}}
   + \sum_i \eta_i^2\, \int_{2M_i}^\infty d\mu\,{c(\mu, M_i)\over M_i^2}\, 
{\mu^2 p^2 \over {p^2 + \mu^2}} \\
&\quad = 4 \cdot {3\over 2} \int_1^\infty {d\nu\over \nu^4}\,
       {1\over \sqrt{\nu^2 - 1}} \left({\nu^2 p^2\over p^2 + 4m^2\nu^2}
         + \sum_i \eta_i^2 {\nu^2 p^2\over p^2 + 4M_i^2\nu^2}\right) \\
 &\quad \to 4\cdot {3\over 2} \int_1^\infty {d\nu\over \nu^4}\,
        {\nu^2\over \sqrt{\nu^2 - 1}} \\
 &\quad = 4\cdot {3\over 2} \cdot \half\cdot B\bigl(\hhalf, 1\bigr) \\
&\quad = 4\cdot {3\over 2} \cdot \half\cdot {\Gamma\bigl(\hhalf\bigr) \,\Gamma \bigl(1\bigr)
  \over \Gamma\bigl({\textstyle{3\over 2}}\bigr)} \\
&\quad = 6,
\end{align*}
where the limit $M_i \to \infty$ and $m\to 0$ has been taken in the 
third line, which makes the Pauli-Villars contributions vanish.  

The contribution (\ref{double}) therefore becomes
\begin{align*} 
 - {1\over (\pi/2)}\, {(ik)^2\over 2!}
   \int {d^2 p\over (2\pi)^2}\, e^{-ip\cdot (x - y)}\,
     {1\over 16}\cdot {\pi\over 3}\cdot 6
= {k^2 \over 8} \, \delta^2 (w - z). 
\end{align*}
This will indeed give the contact term (\ref{contact}) when inserted in 
(\ref{fullcontact}).  

We still need to calculate the contributions of the single contractions
appearing in (\ref{fullcontact}).  
We have
\begin{align*}
&ik\biggl(m^2 \, X(w) \expect{X(w)\,X(z)}
               + \sum_i \eta_i M_i^2\, \chi_i(w) \expect{\chi_i(w)\,\chi_i(z)}\biggr)\\
&\quad = ik\biggl(X(w) \,m^2 \,{1\over 2\pi}\,K_0(m|w - z|)
               + \sum_i \eta_i \,\chi_i(w)\, M_i^2 \,{1\over 2\pi}\,K_0(M_i|w - z|)\biggr)\\
&\quad \to {ik}\,\delta^2(w-z)\sum_i \eta_i \,\chi_i(w)
\end{align*}
in the limit $m\to 0$ and $M_i \to \infty$.
The full expression is thus
\begin{align}
  T_{w\bar w}\, e^{ik\left(X(z) + \sum_i \eta_i \chi_i\right)}
   &= \delta^2 (w - z) \left\{ {k^2 \over 8}
           + {ik}\sum_i \eta_i \,\chi_i(z) \,\right\}
    e^{ik\left(X(z) + \sum_i \eta_i \chi_i(z)\right)} \nonumber\\ 
    &\quad+ \cdots,
\label {fullcontact2}
\end{align}  
which differs from (\ref{fullcontact}) by the presence of 
additional contributions proportional to the Pauli-Villars fields. 
However, as long as no additional insertions are at $z$, the 
matrix elements of these contributions 
go to zero when $M_i\to\infty$, since in this limit
the range of the propagator goes to zero and $\chi_i$ becomes 
non-dynamical.  Modulo this condition, we therefore find  
\begin{align}
  T_{w\bar w}\, e^{ik\left(X(z) + \sum_i \eta_i \chi_i\right)}
   &=  {k^2 \over 8}\, \delta^2 (w - z) \,
    e^{ik\left(X(z) + \sum_i \eta_i \chi_i(z)\right)} + \cdots.
\label {fullcontact3}
\end{align}  
To see that our analysis is consistent with the 
usual operator formalism result, we restate our Ward identity (\ref{vward})
as follows
\begin{align}
 &- {1\over \pi} \int d^2w \,\bar\partial v^w\,\expect{T_{ww}\,
               V_z \cdots} \nonumber 
    - {1\over \pi} \int d^2w\, \partial v^{\bar w}\,\expect{T_{\bar w\bar w}\,
               V_z\cdots} \\
&\qquad =
   \expect{(v^z\partial + v^{\bar z}\bar\partial)\, V_z \cdots}
+ {1\over \pi} \int d^2w \,(\partial v^w + \bar\partial v^{\bar w})\,\expect{T_{w\bar w}\,
               V_z \cdots} \nonumber \\ 
  &\qquad = \expect{(v^z\partial + v^{\bar z}\bar\partial)\, V_z \cdots} + {\alpha' k^2\over 4}\, (\partial_i v^i)\, \expect{V_z\cdots} + \cdots. \label {vward1}
\end{align}
where we used the explicit result
 (\ref{fullcontact3}) for the contact term.  This is the 
familiar anomalous identity from the operator formalism.  

  It is now clear exactly 
how the coordinate dependence arises in the operator formalism.  
In \cite{myself} it was shown that the full renormalized 
energy-momentum $T_{ij} dx^i\otimes dx^j$, including Pauli-Villars 
contributions, is a coordinate-invariant, true
tensor quantity.  But its component $T_{w \bar w}$ of course depends
on the coordinate system, so that by moving the $T_{w \bar w}$
contribution to the right hand side in (\ref{vward1}), one is 
explicitly making both sides of the equation coordinate-dependent.

In the current formalism, the anomalous dimension 
of the insertion is encoded in the contact 
contraction
(\ref{fullcontact3}).  Since $T_{z\bar z}$ is precisely the 
generator of dilations, we have directly related the anomalous 
dimension to scale-dependence, a relationship that is somewhat obscured 
in the usual operator treatment.  We also note that the operator formalism
result is reproduced after ignoring the nonsingular terms, represented by
the dots, in
(\ref{fullcontact3}).  This may be done for the Ward identity but 
needs care in general amplitudes, where these terms may contribute.

From (\ref{vward1}), the precise relationship between our
Pauli-Villars-regulated vertex $V$ and the operator formalism 
vertex operator $\hat V$ is now clear.  
Define 
$$
\hat V = V
$$ 
on the plane with trivial metric, and deform $\hat V$ according to the
transformation law
\begin{align*}
  \delta_v \hat V &\equiv 
v^i\partial_i\hat V 
        + {\alpha' k^2\over 4}\, (\partial_i v^i)\, \hat V
\end{align*}
as we deform only the metric along the flow of a vector field $v$
holomorphic in a neighbourhood of the insertion. 
Then  $\hat V$
coincides with the operator formalism vertex.  

The anomalous term in the transformation
 $\delta_v \hat V$ is an artifact of the coordinate-dependent 
definition of $\hat V$, and is somewhat unnatural in the present formalism.  
On the other hand, the covariantly regularized 
$V$ transforms as a scalar, without this extra term, but requires the 
 nonzero 
contraction with $T_{z\bar z}$ in the Ward identity.  

It should be stressed that both the coordinate-independent 
 $V$ and the coordinate-dependent $\hat V$ are \textit{finite},
renormalized insertions.  The difference between them is a 
finite, but coordinate-dependent quantity.

What about conformal transformations?  Notice that 
(\ref{vward}) is true for arbitrary deformations $v$ 
that goes to a constant at infinity.  Now consider 
a vector field $v$ that is holomorphic
$$
  \partial_{\bar z} v^z = 0 = \partial_z v^{\bar z}
$$
on a disc-shaped neighborhood $D$ of 
$z$.  Since classically 
such a deformation would be a symmetry, one would expect the 
transformation $\mathcal{L}_v V$  to be 
generated by conserved 
charges.  Indeed, as shown in \cite{myself}, away from 
additional insertions the conservation law
$$
\partial_{\bar w} T_{ww} + \partial_w T_{\bar w w} = 0 = \partial_w T_{\bar w \bar w} +  \partial_{\bar w} T_{w w}
$$
holds, so we find, denoting by $\bar D$ the complement of $D$,
\begin{align}
   \expect{(v^z\partial + v^{\bar z}\bar\partial)\, V_z \cdots} 
  = 
    & - {1\over \pi} \int_{\bar D} d^2w \,\expect{\bar\partial\, (v^w\,T_{ww}
      + v^{\bar w}\, T_{\bar w w})\,         V_z \cdots} \nonumber \\
& - {1\over \pi} \int_{\bar D} d^2w \,\expect{\partial\, (v^w\,T_{w\bar w}
      + v^{\bar w}\, T_{\bar w \bar w})\,         V_z \cdots} \nonumber \\
  & 
     - {1\over \pi} \int_D d^2w \,(\partial v^w + \bar\partial v^{\bar w})\,\expect{T_{w\bar w}\,
               V_z \cdots} \nonumber
\end{align}
In the first two terms, since $w$ is outside $D$, the contact term
in the contraction $T_{\bar w w} V_z$ does not contribute.  We obtain
\begin{align}
   \expect{(v^z\partial + v^{\bar z}\bar\partial)\, V_z \cdots} 
  &= 
     {i\over \pi} \oint_{\partial D} dw \,\expect{v^w\,T_{ww}
      \,         V_z \cdots} - {i\over \pi} \oint_{\partial D} d\bar w \,\expect{
      v^{\bar w}\, T_{\bar w \bar w}\,         V_z \cdots} \nonumber \\
  &\quad
     - {1\over \pi} \int_D d^2w \,(\partial v^w + \bar\partial v^{\bar w})\,\expect{T_{w\bar w}\,
               V_z \cdots} \nonumber
\end{align}
The first line represents the contractions with the usual Virasoro charges, 
while the second line is an anomalous term due to quantum breaking of scale
invariance.  As shown in equation (\ref{vward1}), this term is exactly the
anomalous transformation of the operator-formalism vertex.  In the present
covariant formalism, it is more natural, and perhaps less confusing,
 to say that since 
conformal invariance is broken, the conserved 
charges do \textit{not} generate the transformation, with the 
needed correction given by the contact term in the second line.

\section{Ghost path integral}
  
To make this article more self-contained, 
we give a short overview of the ghost component of the path integral, 
which was not considered in the previous sections.  For the purposes
of this calculation, we will not need the 
Pauli-Villars partners for the ghosts.  However, the 
full action does include such additional fields,  and will be derived 
in section \ref{BRSTsection}.  

As in the previous section,
the ghost path integral can be performed by expanding around 
a stationary solution, which can be written on the disc as
\begin{align}
  b_{zz} &= \sum_{m = -\infty}^{-2} b_m z^{-2 - m}, \\
  c^z &= \sum_{m = -\infty}^{1} c_m z^{1 - m},
\end{align}
and similarly for $z \leftrightarrow \bar z$.  The state on the 
boundary can therefore be written as a wave function 
$$
   \psi (\cdots, b_{-4}, b_{-3}, b_{-2}, c_1, c_0, c_{-1}, \cdots)
  \equiv \braket{\cdots, b_{-4}, b_{-3}, b_{-2}, c_1, c_0, c_{-1}, \cdots}{\psi}
$$
on the 
configuration space parameterized by the above expansion coefficients.
This wave function is given by the path integral as \cite{alvarez}
\begin{align*}
  \psi &= \mathcal{Z}\, e^{- \int_D \left(b_{zz}\partial_{\bar z} c^z
     + c^z\partial_{\bar z} b_{zz} \right)
   + \int_{\delta D} dz\, b_{zz}c^z +  (z\leftrightarrow \bar z)} \\
  &= \mathcal{Z}\, e^0 \\
  &= \mathcal Z,
\end{align*}
where the fields $b$ and $c$ in the exponent denote the stationary 
solution.   
The boundary term in the action ensures that the equations of motion 
are satisfied in the presence of the boundary.  
Ignoring the antiholomorphic contribution for now, we 
may write this 
state in terms of the conventionally defined ghost ground state,
given by
$$
  \braket{\cdots, b_{-2}, b_{-1}, c_0, c_{-1}, \cdots}{\downarrow}
  = 1,
$$
by calculating 
\begin{align*}
  \braket{\cdots, b_{-2}, b_{-1}, c_0, c_{-1}, \cdots}{\psi} &=
    \int dc_1\,\braket{b_{-1}}{c_1}\,
\braket{\cdots, b_{-4}, b_{-3}, b_{-2}, c_1, c_0, c_{-1}, \cdots}{\psi} \\
   &= \int dc_1\,e^{i b_{-1} c_1} \,\mathcal{Z} \\
  &= \mathcal{Z}\, b_{-1}  \\
  &= \mathcal{Z}\,\bra{\cdots, b_{-2}, b_{-1}, c_0, c_{-1}, \cdots}b_{-1}
\ket\downarrow,
\end{align*}
where we have used the fact that $b_{-n}$ and $c_n$ are conjugate variables.
We find
$$
  \ket{\psi} = \mathcal{Z}\,b_{-1} \ket{\downarrow},
$$ 
or, including the antiholomorphic contribution,
$$
  \ket{\psi} = \mathcal{Z}\,b_{-1} \bar b_{-1}\ket{\downarrow}.
$$ 
Let us also calculate the path integral over the complement of the 
disc on the sphere.  Taking into account the transformations
\begin{align}
  \partial_z &\to -z^2 \,\partial_z, \\
  dz\otimes dz &\to {1\over z^4}\, dz\otimes dz,
\end{align}
under $z\to 1/z$, the stationary solutions that are regular at 
infinity can be written as
\begin{align}
  b_{zz} &= \sum_{m = -\infty}^{-2} b_m z^{-2 + m}, \\
  c^z &= \sum_{m = -\infty}^{1} c_m z^{1 + m}.
\end{align}
The state on the 
boundary can therefore be written as a wave function 
$$
   \psi (\cdots, b_{4}, b_{3}, b_{2}, c_{-1}, c_0, c_{1}, \cdots)
  \equiv \braket{\psi}{\cdots, b_{4}, b_{3}, b_{2}, c_{-1}, c_0, c_{1}, \cdots}
$$
on the 
configuration space parameterized by the above expansion coefficients.
This wave function is given by the path integral as
\begin{align*}
  \psi &= \mathcal{Z}\, e^{- \int_{\bar D} \left(b_{zz}\partial_{\bar z} c^z
     + c^z\partial_{\bar z} b_{zz} \right)
   - \int_{\delta D} dz\, b_{zz}c^z +  (z\leftrightarrow \bar z)} \\
  &= \mathcal{Z}\, e^0 \\
  &= \mathcal Z.
\end{align*}
The boundary term in the action again ensures the equations of motion 
and cancels the corresponding boundary term on the disc above.  
Ignoring the antiholomorphic contribution for now, we 
may write this 
state in terms of the dual to the conventionally defined ghost ground state,
given by
$$
  \braket{\downarrow}{\cdots, b_{2}, b_{1}, c_0, c_{1}, \cdots}
  = 1,
$$
by calculating 
\begin{align*}
  \braket{\psi}{\cdots, b_{2}, b_{1}, c_0, c_{1}, \cdots} &=
    \int dc_1\,
\braket{\psi}{\cdots, b_{4}, b_{3}, b_{2}, c_{-1}, c_0, c_{1}, \cdots} 
  \,\braket{c_{-1}}{b_{1}}\\
   &= \int dc_{-1}\,e^{i c_{-1} b_1} \,\mathcal{Z} \\
  &= \mathcal{Z}\, b_{1}  \\
  &= \mathcal{Z}\,\bra\downarrow b_{-1}\ket{\cdots, b_{2}, b_{1}, c_0, c_{1}, \cdots}.
\end{align*}
We find
$$
  \bra{\psi} = \mathcal{Z}\,\bra\downarrow b_{1},
$$ 
or, including the antiholomorphic contribution,
$$
  \bra{\psi} = \mathcal{Z}\,\bra{\downarrow} b_{1}\bar b_{1}.
$$ 
We will see later that the fixed vertex operators in string 
theory are accompanied by a factor of 
$$
 c^z c^{\bar z}.
$$  
We 
illustrate the calculation of the path integral over the disc 
with such an insertion 
at the origin
$$
  \psi = \int_D [dc]\, [db]\,c^z(0)\, c^{\bar z}(0)\,
    e^{-S}.    
$$ 
Again, we expand around the stationary solution $c \to c + \tilde c$ to get
\begin{align*}
 \psi =  &c_1 \bar c_1 \, \mathcal Z
    + c_1 \expect {\tilde c^z(0)} 
   + \expect{\tilde c^z(0) \,\tilde c^{\bar z}(0)} \\
 &\qquad = c_1 \bar c_1 \, \mathcal Z,
\end{align*}
since the one-point functions around a stationary background 
are zero, and the two-point function is trivially zero on the plane, 
since there is 
no $c^zc^{\bar z}$ term in the action.  

Again temporarily ignoring the antiholomorphic contribution, we 
calculate
\begin{align*}
  \braket{\cdots, b_{-2}, b_{-1}, c_0, c_{-1}, \cdots}{\psi} &=
    \int dc_1\,\braket{b_{-1}}{c_1}\,
\braket{\cdots, b_{-4}, b_{-3}, b_{-2}, c_1, c_0, c_{-1}, \cdots}{\psi} \\
   &= \int dc_1\,e^{i b_{-1} c_1} \, c_1\,\mathcal{Z} \\
  &= \mathcal{Z}  \\
  &= \mathcal{Z}\,\braket{\cdots, b_{-2}, b_{-1}, c_0, c_{-1}, \cdots}
    \downarrow.
\end{align*}
The result is therefore
$$
  \ket{\psi} = \mathcal{Z} \ket{\downarrow}.
$$ 
Similarly to the above calculation without insertion, 
the state obtained by integrating over the complement of the disc
with an insertion at $\infty$
$$
  \psi = \int_{\bar D} [dc]\, [db]\,c^z(\infty)\, c^{\bar z}(\infty)\,
    e^{-S},
$$ 
is proportional to 
$$
  \bra{\downarrow}.
$$
The path integral over the full sphere with the two insertions is 
then given by integrating these states over the configuration space on the 
boundary
\begin{align*}
  \braket{\downarrow}\downarrow
   &= \int \left( \cdots db_{-2}\, db_{-1}\, dc_0\, dc_{-1}\cdots\right)
       \braket{\downarrow}{\cdots, b_{-2}, b_{-1}, c_0, c_{-1}, \cdots} 
\times \\
 &\qquad \qquad \qquad\qquad \qquad \qquad\qquad \qquad \times
        \braket{\cdots, b_{-2}, b_{-1}, c_0, c_{-1}, \cdots}{\downarrow} \\        
   &= \int \left( \cdots db_{-2}\, db_{-1}\, dc_0\, dc_{-1}\cdots\right)
       \braket{\downarrow}{\cdots, b_{2}, b_{1}, c_0, c_{1}, \cdots}
        \braket{c_1}{b_{-1}}\braket{c_2}{b_{-2}} \cdots \\
   &\qquad \qquad \qquad\qquad \qquad \qquad\qquad \qquad \times
       \braket{\cdots, b_{-2}, b_{-1}, c_0, c_{-1}, \cdots}{\downarrow} \\
   &= \int \left( \cdots db_{-2}\, db_{-1}\, dc_0\, dc_{-1}\cdots\right)
          (1)\, e^{ic_1b_1} e^{ic_2 b_2} \cdots \,(1) \\
   &= 0,
\end{align*}
due to the integral over $c_0$.  We see that the path integral on the sphere 
with two fixed vertex insertions vanishes.  It is clear that an extra 
insertion at a third position, proportional to, for example,
$$c^z(1)\,\bar c^z(1),$$
will be sufficient to provide an extra factor $c_0 \bar c_0$ to make 
the path integral non-vanishing.  In other words, the first 
non-vanishing string amplitude will be the three-point function.

\section{Background-independent measures}
\label{measuresection}

In \cite{myself}, the background dependence of the path integral
measure for a scalar field $\phi$ was studied.  The Fujikawa 
approach \cite{fujikawa1, nakahara}  that we followed there depended 
on the existence of an a priori background metric $g^{ij}$
for the construction of the measure, denoted $[d\phi]_{g}$.
Varying $g$ caused a variation in $[d\phi]_{g}$ that 
could be related to the conformal anomaly. 

We then introduced a set of Pauli-Villars auxiliary fields $\chi$ with 
statistics opposite to that of $\phi$.  Due to the 
properties of Grassmann fields, the corresponding measure $[d\chi]_{g}$
transformed with opposite sign to $[d\phi]_{g}$ under a variation 
of $g$.  As a result, the combined measure
$$
  [d\phi]_{g}[d\chi]_{g} = [d\phi]_{g'}[d\chi]_{g'}
    \equiv [d\phi]\, [d\chi]
$$
was background-independent.

Something similar can be done for arbitrary tensor fields by
including Pauli-Villars partners in this way.  This inclusion
of Pauli-Villars fields to simplify the transformation properties
of the measure
differs from the approach to string quantization 
followed by Fujikawa in \cite{fujikawa2}, although individual 
factors contributing to the measure are defined in essentially the 
same way.  
 
We start with the example of a vector field.
The Fujikawa measure for a vector field can be constructed as
follows.  We make the space of vector fields on the plane 
into an inner product space by defining 
$$
  \braket{v}{w}_{g} \equiv \int \sqrt{g}\, g_{kl}\, v^k w^l.
$$
We then consider an orthonormal basis $\phi^k_n(x)\,\partial_k$
of vector fields on the plane, where $n$ is an abstract index 
ranging, in the two-dimensional case, over $\mathbf{N}\times\{1, 2\}$. 
In other words, 
\begin{align}
  \int d^2x\, \sqrt {g}\, g_{kl}\,\phi^k_n \phi^l_m = \delta_{mn}.
\label {ortho}
\end{align}
Parameterizing $v^k$ in terms of a countable set of coordinates
$a_n$ via the basis expansion
$$
  v^k(x) = \sum_n a_n\,\phi^k_n(x),
$$
we define the measure via the semi-infinite differential form
$$
  [dv]_{g} \equiv \bigwedge_{n} da_n
$$
on the infinite-dimensional space of fields.  Since different 
orthonormal bases are related by unitary transformations 
that have Jacobian $1$, the differential form is in fact independent
of the choice of basis.

Since the basis $\phi^k_n$ depends by construction on $g^{ij}$,
the measure is background-dependent.  We may calculate the precise
dependence as follows.  First, note that we can project
$$
  a_n = \int \sqrt g\, g_{kl} \,\phi^l_n v^k,
$$
so that 
$$
  \delta_{g} a_n = \int \sqrt {g} \left(\half\, g^{ab}\delta g_{ab}\, g_{kl}
       + \delta g_{kl} \right)  \phi^l_n v^k
    + \int \sqrt {g} \, g_{kl} \, (\delta_{g} \phi^l_n)\, v^k.
$$
The variation $\delta_{g} \phi^l_n$ of the orthonormal basis  is not 
unique, but may be determined up to a unitary transformation,
which does not affect the measure, by differentiating (\ref{ortho})
to find
$$
  0 = \int \sqrt {g} \left( \half\, g^{ab}\,\delta g_{ab}\, g_{kl}
          + \delta g_{kl} \right) \phi^k_n \phi^l_m
   + \int \sqrt {g} \, g_{kl} \left((\delta\phi^k_n) \,\phi^l_m + \phi^k_n\, (\delta\phi^l_m)
       \right)
$$
A suitable choice that satisfies this is 
$$
  \delta \phi^k_n = - {1\over 4}\, g^{ab}\,\delta g_{ab}\,\phi^k_n 
   - \half \, g^{kp} \,\delta g_{pq}\, \phi^q_n.
$$
Inserting in the above,  we find 
\begin{align*}
  \delta_{g} a_n &= \int \sqrt g \left({1\over 4}\, g^{ab}\,\delta g_{ab}\, g_{kl} 
       + \half\, \delta g_{kl}\right) \phi^l_n v^k \\
    &= \sum_n a_m \int \sqrt g \left({1\over 4}\, g^{ab}\,\delta g_{ab}\, g_{kl}
       + \half\, \delta g_{kl}\right) \phi^l_n \phi^k_m \\
    &\equiv \sum_n C_{nm}\, a_m,
\end{align*}
so that the measure transforms as 
$$
  \delta_g [dv]_g =
  \delta_g  \bigwedge_{n} da_n 
    = (\mathrm{Tr}\, C)\, \bigwedge_{n} da_n,   
$$
where
\begin{align}
  \mathrm{Tr}\, C = \sum_m C_{mm}  &= \int d^2 x \sqrt g
    \left({1\over 4}\, g^{ab}\,\delta g_{ab} \, g_{kl}
       + \half\, \delta g_{kl}\right)
    \sum_m \phi_m^k(x)\,\phi_m^l(x) 
\label{anom}
\end{align}
The sum over $m$ does not converge and has to be regulated.  This
may be done, for example, by inserting a heat kernel regulator,
considering instead
$$
  \sum_m \phi_m^k(x)\,e^{\epsilon \Delta_g} \,\phi_m^l(x),
$$
for small $\epsilon$, 
where $\Delta_g \equiv \nabla^a\nabla_a$ is 
the Laplacian on the space of vector fields.  These heat kernels
may be calculated as, for example, in \cite{myself}, and typically lead to 
a divergence of order $1/\epsilon$, which may be canceled by a 
counterterm in the action, and a finite contribution depending on the 
curvature that cannot be canceled by a counterterm.  
This contribution can be related to the conformal anomaly.  

However, we can cancel the background dependence of the 
measure
by including Pauli-Villars auxiliary vector fields.  As in 
\cite{myself}, we introduce a set of these, consisting of 
real commuting fields 
$\nu^k_i$, each with statistics $c_i=1$, and complex Grassmann fields $
\nu_i^k$, $\bar\nu_i^k$, each with statistics $c_i = -2$, satisfying 
$$
  \sum_i c_i = -1.  
$$
The Grassmann fields have to be taken complex, otherwise the 
mass term $$M^2 \,g_{kl}\,\bar\nu^k \nu^l$$ would vanish for them.  
Due to the transformation properties of Grassmann integrals, we find,
denoting the combined measure of all the Pauli-Villars fields by
$[d\nu]_g$, that
$$
  \delta_g [d\nu]_g =   \left(\sum_i c_i\right) (\mathrm{Tr}\, C)\,
   [d\nu]_g  =  -  (\mathrm{Tr}\, C)\,
   [d\nu]_g,
$$
where the sign is \textit{opposite} to that of the original fields.  
Therefore 
$$
  \delta_g \left ( [dv]_g \wedge [d\nu]_g\right) = 
  \biggl((\mathrm{Tr}\, C) - (\mathrm{Tr}\, C)\biggr) [dv]_g \wedge [d\nu]_g
  =
  0
$$
with \textit{any} reasonable regularization.   
In other words, the combined measure is background-independent, 
independently of the regularization scheme. 

Once we include the Pauli-Villars fields, the measure does not 
transform anomalously.  So what happened to the 
conformal anomaly?  The burden of encoding this is now transferred to
the Pauli-Villars action.  To understand this, notice that 
the anomalous contribution to the transformation of the
original measure, given by 
$$
\int d^2 x \sqrt g
    \left({1\over 4}\, g^{ab}\,\delta g_{ab} \, g_{kl}
     +  \half\, \delta g_{kl}\right)
    \sum_m \phi_m^k(x)\,\phi_m^l(x), 
$$
is just the energy-momentum contribution we would obtain from 
the mass terms of the Pauli-Villars fields $\nu$
\begin{align*}
 \delta_g \int [dv]\wedge[d\nu]\, e^{-S} 
& =  
   -{1\over 4\pi}\int d^2 x \, \sqrt g\,\delta g_{kl} \expect{T_\nu^{kl}} \\
& \to 
 - \delta_g  \int d^2 x\, \sqrt g\,\half\, M^2\, g_{kl} \expect{\bar\nu^k(x)\,\nu^l(x)} 
\end{align*}
in the limit as the mass tends to infinity, since
then the self-contraction becomes  
$$
  \expect{\bar\nu^k(x)\,\nu^l(x)} \to - {2\over M^2} \sum_m \phi_m^k(x)\,\phi_m^l(x) 
$$
for the Grassmann-valued  $(\bar \nu_i, \nu_i)$, and
$$
  \expect{\bar\nu^k(x)\,\nu^l(x)} \to {1\over M^2} \sum_m \phi_m^k(x)\,\phi_m^l(x) 
$$
for the commuting partners $\bar \nu_i = \nu_i$.  Since $\sum_i c_i = -1$, 
it follows that the mass terms needed to make the Pauli-Villars fields
non-dynamic will now contribute the anomaly.  

A similar analysis can be done for other tensor fields.  For example,
consider the Grassmann-valued ghost vector field $c = c^i\partial_i$.  
This is a vector field and, as above, we define the measure as
$$
  dc_1 \wedge dc_2 \wedge \cdots,
$$
where $c_n$ are the expansion coefficients with respect to the 
above orthonormal
basis $\phi^k_n$.

The only difference from the above analysis appears in the mass term.
A non-vanishing covariant scalar mass term for a ghost $c^i$ would be
$$
  \half\,m^2\left(\sqrt g\, dx^1\wedge dx^2\right) (c, c) = 
  \half\,m^2\,\sqrt g\, \epsilon_{ij}\, c^i c^j,
$$
and this is the form of the mass term for the Grassmann
subset of the ghost Pauli-Villars partners.
The bosonic Pauli-Villars partners, which have to be 
complex fields for this expression to be non-vanishing,
have mass terms
$$
 \half\, M^2\left(\sqrt g\, dx^1\wedge dx^2\right) (\bar \gamma, \gamma),
$$
Note, however, that these terms are not positive definite.  As a result,
their Euclidean path integral $\int [d\gamma]\,e^{-S}$ does not exist.  
However, a consistent non-perturbative construction
of the real time path integral $\int [d\gamma]\,e^{iS}$ can be done
\cite{myselffuture}, 
which justifies the formal manipulations carried out on this object
in the Euclidean picture by
treating it as if it were convergent.

The anomalous variation of the ghost measure is, by construction, opposite to
that of the vector measure.  When the Pauli-Villars fields are included,
the total measure is again invariant, and 
the anomaly that would otherwise have come from the 
measure will again have 
to be encoded in the Pauli-Villars energy momentum tensor.  

This means that a ghost anomaly, calculated for the 
Pauli-Villars action,  should be opposite to a vector anomaly calculated 
in the first part of this section,
even though the mass terms have a different form.  To see that this is true,
we may use conformal coordinates without loss of generality, 
since nothing in the above construction
depends on the coordinate choice. 
The mass contribution for a vector partner is
$$
   \half\, M^2 \int d^2x\,\sqrt g\, g_{z\bar z}\, \bar\nu^{\bar z} \nu^{z}
   = \half\,M^2 \int d^2x\,g_{z\bar z}^2 \,\bar\nu^{\bar z} \nu^{z},
$$ 
and for a ghost partner it is
$$
\half\,M^2 \int d^2x\,\sqrt g\,\sqrt g \,\epsilon_{z\bar z}\,\bar\gamma^{\bar z} \gamma^z
  = \half\,M^2 g_{z\bar z}^2 \bar\gamma^{\bar z} \gamma^z,
$$
while an arbitrary variation of the metric (not necessarily conformal)
gives for a vector partner
\begin{align*}
  \half\, M^2 \delta\left(\sqrt g\, g_{z\bar z}\right) \expect{\bar\nu^{\bar z} \nu^{z}}
   &= \half\, M^2 \left(\half \sqrt g\left(g^{kl}\,\delta g_{kl}\right)
   g_{z\bar z} + \sqrt g\, \delta g_{z\bar z}\right) \expect{\bar\nu^{\bar z} \nu^{z}}\\
   &=  M^2 \, g_{z\bar z} \,\delta g_{z\bar z}
\expect{\bar\nu^{\bar z} \nu^{z}}
\end{align*}
and for a ghost partner
\begin{align*}
\half\,M^2 \,\delta g\, \epsilon_{z\bar z}\expect{\bar\gamma^{\bar z} \gamma^z}
&= \half\,M^2 \,\delta \,(g_{z\bar z}^2)\, \epsilon_{z\bar z}\expect{\bar\gamma^{\bar z} \gamma^z} \\
&=  M^2 \, g_{z\bar z} \,\delta g_{z\bar z}
\expect{\bar\gamma^{\bar z} \gamma^z} 
\end{align*}
Since, in the limit of large $M$, the mass terms become dominant in the action,
and since these coincide for the two cases above, we find the same 
contribution except for a sign.  In other words, the ghost anomaly,
as  encoded in the unusual
ghost Pauli-Villars mass term, is indeed
opposite to the vector anomaly.

The measure on the space of metrics $g^{ij}$ is particularly important.
We construct the background-independent 
path integral measure over $g^{ij}$ as above as follows.
Choose a background metric $\bar g^{ij}$, and let $[dg]_{\bar g}$ be
the Fujikawa measure on the space of metrics with respect to $\bar g^{ij}$.
In other words, we 
expand, similar to the above, in a
basis on the space of symmetric tensors $g^{ij}$ orthonormalized in 
the obvious inner product with respect to $\bar g^{ij}$, and define
the measure, as above, as a differential form 
in terms of the expansion coefficients.
This measure depends on $\bar g^{ij}$, and is therefore not yet suitable
for a background-independent theory.  To obtain a background-independent 
measure, introduce a set of Pauli-Villars auxiliary fields, denoted
$\{\gamma^{ij}, \bar\gamma^{ij}\}$ that are either real commuting
with $\bar \gamma = \gamma$
and with statistics $1$, or 
complex Grassmann with statistics $-2$, whose combined statistics 
is equal to $-1$ to cancel that of the original $g^{ij}$  
(the Grassmann $\gamma$ are symmetric, complex-valued
matrices, not hermitian matrices).  As above, the combined
measure is then independent of $\bar g$, independently of which 
regularization we choose.  In particular,
$$
  [dg]_{\bar g}\wedge[d\gamma]_{\bar g}
    =  [dg]_{g}\wedge [d\gamma]_{g},
$$
so that the combined measure is background independent. 

A slight complication that has to be kept in mind is that the range 
of an integration over the metrics is not unconstrained, as was
the case for scalar or vector fields, since we 
have to ensure that the metric does not become degenerate or 
change signature.  
In the two-dimensional case, we will manage to avoid 
the need for integrating over the metrics, so
this issue will not affect the analysis that
follows, but it should be kept in mind in higher dimensions.

\section{Physical state conditions}
\label{physicalsection}

In the operator formalism, the physical state condition on the 
vertex operators $\hat V$ are related to their anomalous dimension.  
In the current formalism, the Ward identity (\ref{vward}) for the insertion
$V = \exp\, ({ik\,(X + \sum \eta_i\chi_i)})$ is that of a scalar, and the
anomaly is
encoded in the non-vanishing contact term matrix elements 
of $T_{w\bar w}\,V_z$.  
 Given this difference, it is very instructive to 
derive the physical state condition
on $V$ in the current coordinate-independent formalism.  

Throughout this section we will use the covariant form of the path integral 
derived later in (\ref{covBRST}).

Consider a path integral with $n$ vertex insertions, where the path 
integral includes an integral over the positions $x_i$, $i = 1, \dots, n$ 
of the insertions.
By a gauge transformation consisting of diffeomorphisms and 
Weyl transformations, we can bring the metric to a chosen form $\hat g$,
modulo possible 
global obstructions that do not occur on the plane or the sphere.
By further conformal transformations (which are compositions 
of diffeomorphisms and conformal transformations leaving the metric
invariant), we may 
fix the positions of $m\le n$ of the $x_i$ to be at chosen 
points $\hat x_i$, $i = 1,\dots , m$, where the value of $m$ depends on the 
topology.  On the plane, $m$ is one,
since $v$ may go to a constant at infinity, 
whereas on the sphere, a M\"obius transformation may be used to 
fix $m = 3$ positions in this manner.  

Consider then a family of gauge fixing functions
$$\delta (g^{ij} - \hat g^{ij})\, \delta (x_1 - \hat x_1) \, \cdots\,
 \delta (x_m - \hat x_m),$$ 
indexed by 
$(\hat g^{ij}, \hat x_1, \dots, \hat x_m)$.  
Here $g^{ij}$ and $x_i$ are the metric
and positions over which we integrate in the path integral, and 
$\hat g^{ij}$ and $\hat x_i$ are a fixed metric and fixed positions.  
Gauge fixing will discussed in great detail in section \ref{BRSTsection} and 
in the BRST context, but for the purposes of this section we 
will only need 
the familiar covariant form of the full action and the 
resulting form (\ref{insertion}) of the fixed vertices.  

Any physical quantity has to 
be independent of the hatted quantities determining the gauge fixing function.
We will show how this requirement leads to the physical state 
conditions on the vertex insertions, and later relate it to a
covariant version of BRST invariance in the quantum field theory.

Consider a variation $(\delta \hat g^{ij}, \delta \hat x_1.\dots)$ of the fixed metric
and positions under a gauge transformation consisting of a combined 
reparametrization $v^i\partial_i$ and Weyl transformation.

The next step is to write the gauge fixing function $\delta (g^{ij} - \hat g^{ij})$ in a more convenient form.  To do this, we introduce a Langrange
multiplier field $B_{ij}$, together with its set of Pauli-Villars partners,
conveniently denoted by
$\{\beta_{ij}, \bar \beta_{ij}\}$.  As above, their combined measure
is background-independent
$$
[dB]_{\bar g}\wedge[d\beta]_{\bar g}
    =  [dB]_{g}\wedge[d\beta]_{g}
$$
and we may write
\begin{align*}
  \int [dg]_{\bar g} \,\delta(g^{ij} - \hat g^{ij})
   &= \int [dg]_{\bar g}\wedge [dB]_{\bar g}\wedge[d\beta]_{\bar g}
       \wedge [d\gamma]_{\bar g} \\
 &\qquad \times
      \, \exp i \left\{\left\langle B,\, (g - \hat g) \right\rangle_{\bar g} + \half\left(\left\langle \beta,\, \gamma \right\rangle_{\bar g}
   + \left\langle \bar\gamma,\, \bar\beta \right\rangle_{\bar g}\right)\right\} \\
   &= \int [dg]_{\bar g} 
\wedge [dB]_{\bar g}\wedge[d\beta]_{\bar g}
       \wedge [d\gamma]_{\bar g} \\
 &\qquad \times\exp i \int d^2x \,\sqrt{\bar g}\, 
  \left\{ B_{ij}\, (g^{ij} - \hat g^{ij}) 
+ \half\left(\bar\beta_{ij}\gamma^{ij} 
   + \bar\gamma^{ij}\, \beta_{ij} \right)
 \right\},
\end{align*}
since the integral over $\beta$ and $\gamma$ gives $1$. 
This formula is an easy consequence of the definition of the Fujikawa 
measures in terms of orthonormalized modes with respect to $\bar g^{ij}$.  
Note that the background metric $\bar g$ is different from the 
metric $\hat g$ appearing in the gauge fixing function.  Since the
result of integrating over $\beta$ and $\gamma$ is $1$ independently
of $\bar g$, it follows that the remaining expression
$$
  \int [dg]_{\bar g} \,[dB]_{\bar g}\, \exp i \int d^2 x\, \sqrt{\bar g}\,
  B_{ij}\, (g^{ij} - \hat g^{ij})
$$
is itself independent of $\bar g$.  It is indeed not difficult to 
show this explicitly by demonstrating that the variation of the 
measure $[dg]_{\bar g} [dB]_{\bar g}$ under a change of $\bar g$ cancels the variation
of the integrand.  Indeed, one finds
\begin{align*}
  &\delta_{\bar g} \int 
    [dg]_{\bar g} [dB]_{\bar g}\, \exp i \int d^2 x\, \sqrt{\bar g}\,
   B_{ij}\, (g^{ij} - \hat g^{ij}) \\
&\qquad = 
   \left( \mathrm{Tr}\, C +  \mathrm{Tr}\, C - 2\, \mathrm{Tr}\, C\right)
    \int 
    [dg]_{\bar g} [dB]_{\bar g}\, \exp i \int d^2 x\, \sqrt{\bar g}\,
  B_{ij}\, (g^{ij} - \hat g^{ij}) \\
&\qquad = 0.
\end{align*}
The first two terms come from the variations of the two measures, as in the 
previous section, while
the third comes from evaluating the expectation value obtained
when varying the integrand.
Here
$$
  C_{mn} \equiv  \half \int\delta 
    \left(\sqrt{\bar g}\,
     \bar g_{ik}\,\bar g_{jl}\right) \phi^{kl}_m \phi^{ij}_n,
$$
for a $\bar g$-orthonormal basis $\phi^{kl}_m$ of symmetric tensors.  

However, it should be noted that this argument only works for 
the vacuum amplitude.  Insertions containing $g^{ij}$ in the amplitude
will in general cause the invariance to break down due to 
additional contractions with $B^{ij}$.  
We therefore have to keep the $\bar g$ dependence
in the action separate.  In particular, we \textit{cannot} use this to set
$\bar g^{ij} = g^{ij}$ for general amplitudes.  

The full background-independent measure is given by
\begin{align*}
 d\mu_{\bar g} &\equiv [dg]_{\bar g}\wedge [d\gamma]_{\bar g}\wedge [dB]_{\bar g}\wedge[d\beta]_{\bar g}\wedge [dX]_{\bar g} \wedge [d\chi]_{\bar g}
\wedge \mathrm{ghost} \\
 &= d\mu_g \\
 &\equiv d\mu
\end{align*}
where the ghost contribution includes the Pauli-Villars partners for 
each ghost, to be considered more carefully in section \ref{BRSTsection}.

The path integral will be independent of the gauge fixing function if
\begin{align*}
  0 &=\left(\delta_{\hat g} + \delta_{\hat x_i}\right) \int d\mu \,\left(\sqrt g\,
\epsilon_{kl}\,c^kc^l\, V\right)(\hat x_1) \cdots e^{-S + i\int d^2x\,\left\{B_{ij}\, (g^{ij} - \hat g^{ij}) 
+ \half\left(\bar\beta_{ij}\gamma^{ij} 
   + \bar\gamma^{ij}\, \beta_{ij} \right)
 \right\} }
 \\
      &= -i\int d\mu \,\tilde V(\hat x_1)\cdots  e^{-\tilde S}\, \int d^2x\,\sqrt{\bar g}\,
  B_{ij}\,\delta \hat g^{ij} \\
      &\quad   - v^i \hat\partial_i \expect {\tilde V (\hat x_1)\cdots}_{\tilde S},
\end{align*}
where
\begin{align*}
  \tilde S &\equiv S - i\int d^2x\, \sqrt{\bar g}\, 
\left\{B_{ij}\, (g^{ij} - \hat g^{ij}) 
+ \half\left(\bar\beta_{ij}\gamma^{ij} 
   + \bar\gamma^{ij}\, \beta_{ij} \right)
 \right\}, \\
\tilde V &\equiv \sqrt g\,
\epsilon_{kl}\,c^kc^l\, V
\end{align*}
and the ghost factors accompanying  the fixed vertices come from the 
Faddeev-Popov determinant and are derived in section \ref{BRSTsection}.
The action $\tilde S$ is in fact derived in detail later and is 
given by equation (\ref{covBRST}).
If $V$ is a scalar, then $\tilde V$ will be a 
coordinate-independent scalar quantity, 
since the prefactor is simply the application 
$$
  \left(\sqrt{g}\,dx^1\wedge dx^2\right) (c, c)
$$
of a coordinate-independent bilinear density to the pair $(c, c)$.

Remembering the definition of the energy momentum-tensor, 
we have
$$
  {\delta\over \delta g^{ij}}\, e^{-\tilde S}
    = i \sqrt{\bar g}\, B_{ij} \,e^{-\tilde S} 
                  + {1\over 4\pi}\,\sqrt{g}\, T_{ij}\, e^{-\tilde S},
$$
where $T_{ij}$ denotes the part of the energy-momentum tensor of 
$\tilde S$ independent 
of $B$.  Notice that the terms in 
$\beta$ and $\gamma$ are independent of $g^{ij}$ and therefore do not
contribute to the energy-momentum tensor.   

 We obtain the condition 
\begin{align*}
  0 &= \int d\mu \, \tilde V(\hat x) \cdots \int d^2x\left(
   - {\delta\over \delta g^{ij}(x)}  
    + {1\over 4\pi}\,\sqrt{g}\, T_{ij}
   \right) e^{-\tilde S} 
    \,\delta \hat g^{ij} - v^i \hat\partial_i \expect {\tilde V\cdots}_{\tilde S}\\
   &= \int d^2x\int d\mu\, e^{-\tilde S} 
       \left({\delta\over \delta g^{ij}(x)}\, \tilde V(\hat x)
         + {1\over 4\pi}\,\sqrt{g}\,T_{ij}(x)\,\tilde V(\hat x) \right) \delta \hat g^{ij}(x) \cdots - v^i \hat\partial_i \expect {\tilde V\cdots}_{\tilde S}\\
   &= \expect{{\delta \tilde V\over \delta g^{ij}}\cdots}_{\tilde S}
   \, \delta \hat g^{ij}(\hat x) + {1\over 4\pi} \int d^2x
        \expect {\sqrt{g}\, T_{ij}(x)\, \tilde V(\hat x)\cdots}_{\tilde S}\delta \hat g^{ij}(x)
    - v^i \hat\partial_i \expect {\tilde V\cdots}_{\tilde S} \\
 &= 
\expect{{\delta \tilde V^{\hat g}\over \delta \hat g^{ij}}\cdots}_{S(\hat g)}
   \, \delta \hat g^{ij}(\hat x)
   +{1\over 4\pi} \int d^2x\,\sqrt{\hat g} \expect { T_{ij}^{\hat g}(x)\, \tilde V^{\hat g}(\hat x)\cdots}_{S(\hat g)} \delta \hat g^{ij}(x)
    - v^i \hat\partial_i \expect {\tilde V^{\hat g}\cdots}_{S(\hat g)}
\end{align*}
where we have used translation invariance of the path integral 
measure with respect to $g^{ij}$ to perform a partial integration
in the second step.  
In the last step, we have performed the
path integral over 
$[dg]\wedge [d\gamma]\wedge [dB]\wedge[d\beta]
$,
which fixes $g = \hat g$, 
and we are left with the path integral 
$$
\expect{\cdots}_{S(\hat g)}\equiv
\int
[dX]_{\hat g} \wedge [d\chi]_{\hat g}
\wedge (\mathrm{ghost})\, e^{-S({\hat g})}\, (\cdots),
$$
where the measure is in fact independent of $\hat g$.
We will henceforth drop the $S(\hat g)$ subscripts 
on expectation values.  

But if we decompose the variation of $\hat g^{ij}$ in the gauge directions 
as
$$
  \delta \hat g^{ij} = -\nabla^i v^j - \nabla^j v^i - 2\, \delta\omega\,g^{ij}, 
$$
then the $v$-dependent part of the above condition is exactly zero by 
the Ward identity (\ref{ddlambdaV}, \ref{lvV}), and all that remains
is the $\delta\omega$-dependent part.  The physical state condition 
becomes simply
\begin{align}
 0 = \expect{{\delta \tilde V\over \delta \hat g^{ij}}\,\hat g^{ij}\cdots}
   \delta \omega(\hat x)
   +{1\over 4\pi} \int d^2x\,\sqrt{\hat g} \expect { T_i^{\phantom{i}i}(x)\, \tilde V(\hat x)\cdots} \delta \omega(x) \label{Vcondition}  
\end{align}
This condition has an intuitively reasonable interpretation.  It 
requires that the path integral with insertion 
be invariant under a local Weyl rescaling,
generated by $T_i^{\phantom{i}i}$, of the metric $\hat g^{ij}$ that  
indexes the choice of gauge-fixing function.  It is also interesting
to see that invariance with respect to
diffeomorphisms is automatic, and puts no further constraints on 
$\tilde V$.

The simplest scalar insertion satisfying this gauge invariance condition,
for certain values of $k$, is the tachyon
$$
 \tilde V = \sqrt{g} \,\epsilon_{kl} \,c^k c^l\, e^{ik\tilde X}
   \to c^z c^{\bar z}e^{ik\tilde X},
$$
where $\tilde X$ stands for the combination
$$
  \tilde X \equiv X + \sum_i \eta_i \chi_i.
$$
No regularization is needed for the ghost prefactor,
since the self-contraction of $c^k c^l$ is trivially zero due to the 
absence of any $c^k c^l$ term in the massless action for the ghosts. 

For simplicity, we will concentrate on variations of the flat metric on 
the plane $\hat g^{z\bar z} = 2$, $\hat g^{zz} = 0 = \hat g^{\bar z \bar z}$.
Inserting the vertex into the condition (\ref{Vcondition}), and using 
$$
  {\partial \sqrt g \over \partial g^{ij}} = -\half \, \sqrt g\, g_{ij},
$$
we obtain
$$
  -  \delta\omega(z) \expect{\tilde V(z)\cdots} + {1\over 4\pi}
\int d^2w\,\expect {( T_{w\bar w} + T_{\bar w w})\, c^z c^{\bar z}e^{ik\tilde X(z)}\cdots } \, (2\,\delta\omega(w)) = 0.
$$
We have already calculated the contraction of $T_{w\bar w}$ with
$e^{ik\tilde X}$ in section \ref{wardsection}.  We still have to consider 
the contribution
\begin{align*}
  &\expect{T_{w\bar w}\,c^z c^{\bar z}}\, e^{ik\tilde X}.
\end{align*}
Since the $\{b, c\}$ theory is massless, so that
$T_{w\bar w}(b, c) = 0$, and no
Pauli-Villars contractions are involved here, this expression
trivially vanishes.  

Using the explicit result from section \ref{wardsection},
\begin{align}
  T_{w\bar w}\,
               e^{ik\tilde X_z}   = {\pi}\,{\alpha' k^2\over 4} \,\delta^2(w - z) 
   \,e^{ik\tilde X_z} + \cdots,  \label{tzbarzv} 
\end{align}
we find
$$
  \left(\half - {1\over 2\pi}\,\pi\, {\alpha'k^2\over 4}\right) \expect{\tilde V\cdots} = 0 
$$
so that we obtain the usual tachyon mass-shell condition
$$
   k^2 =  {4\over \alpha'}.
$$
Next we  consider the graviton vertex, given by the 
worldsheet scalar 
\begin{align*}
  \tilde V &= {1\over 4} \, \sqrt g\, \epsilon_{kl}c^k c^l\, g^{ab}\, e^{ij}
    \,\partial_a \tilde X_i \partial_b \tilde X_j\, e^{ik\tilde X}  \\
   &\to c^z c^{\bar z}\, e^{ij}\, \partial_z \tilde X_i\, \partial_{\bar z} \tilde X_j\,
      e^{ik\tilde X}
\end{align*}
where $e^{ij}$ is a spacetime polarization.  
To evaluate the first terms in the condition (\ref{Vcondition}),
we calculate
$$
  g^{ij}\,{\partial \sqrt g\, g^{ab}\over g{ij}} 
   = \sqrt g \left( -\half\, g_{ij}\,g^{ab} + \delta_i^a \delta_j^b\right) g^{ij}  = 0,
$$
so that the gauge invariance condition becomes simply
\begin{align}
  {1\over 4\pi}\int d^2w\,\expect {(T_{w\bar w} + T_{\bar w w}) \, \tilde V(z) \cdots}
     (2\, \delta\omega (w)) = 0.
  \label{graviton}
\end{align}
To verify  the consistency of our covariant approach, we will now 
calculate the contact terms in this condition from first principles.
The calculation is complicated, and this is not the recommended
way of doing things.  In the next section, we will 
discuss the relationship between the above physical state condition 
and the familiar Virasoro conditions.  The latter may be 
computed using familiar methods in the literature, and the results 
may then be  used to infer the contact terms indirectly.

With this in mind, let us return to the 
direct calculation.  In holomorphic coordinates, where 
$$
  \tilde V(z)\equiv  c^z c^{\bar z}\, \partial_z\left(X(z) + \sum_i \eta_i \chi_i(z)\right)
     \partial_{\bar z}\left(X(z) + \sum_i \eta_i \chi_i(z)\right) e^{ik\left(X(z) + \sum \eta_i \chi_i(z)\right)}
$$
the graviton condition becomes
\begin{align}
  &T_{w\bar w}\, \tilde V(z)\nonumber \\
   &\quad= {\pi\over 2} \left(m^2 X^2 + \sum_j M_j^2 \chi_j^2\right)V(z)
  \nonumber\\
  &\quad= {\pi\over 2}\,{(ik)^2\over 2!}
     \biggl( m^2\expect{X^2(w)\,X^2(z)} 
           + \sum_i \eta_i^2 M_i^2 \expect{\chi_i^2(w) \, \chi_i^2(z)}\biggr)
   V(z)\nonumber \\
& \qquad  + {\pi \over 2}\,ik\biggl(m^2 \,\expect{X^2(w)\,(\partial_z X)\, X(z)}
               + \sum_i \eta_i^2 M_i^2\, \expect{\chi_i^2(w)\,
     \,(\partial_z\chi_i)\, \chi_i(z)}\biggr) \nonumber\\
 &\qquad\quad\times \partial_{\bar z}\left(X(z) + \sum_i \eta_i \chi_i(z)\right) e^{ik\left(X(z) + \sum_i \eta_i \chi_i\right)} \nonumber \\
 &\qquad + (z \leftrightarrow \bar z) \nonumber \\
 &\qquad + \cdots. \label{tgraviton}
\end{align}
In the above, contributions from single contractions are absent, since 
these are either proportional to 
$$
  {m^2 \over z - w}\, \partial_{\bar z}X \to 0,
$$
as $m\to 0$ for the matter fields, or proportional to
$$
 M_i^2 \,\partial_{z} K_0 (M_i|w-z|) \, \partial_{\bar z}\chi_i
   \to 2\pi\,\partial_{z} \delta^2(w - z) \, \partial_{\bar z}\chi_i,
$$
as $M_i \to \infty$, but matrix elements of $\chi_i (z)$ vanish in this 
limit
as long as no other insertions approach the point $z$.  
We have also dropped double contractions of the 
form $m^2\expect {X^2(w)\, \partial_z X\partial_{\bar z} X}$ proportional to 
$$
   m^2 \, {1 \over w - z}\, \, {1 \over \bar w - \bar z} \to 0,
$$
and the corresponding Pauli-Villars contributions, proportional to 
$$
  M_i^2 \, \partial_z K_0(M_i |w - z|)\, \partial_{\bar z}
    K_0(M_i |w - z|) \to (2\pi)^2\,\partial_z \delta^2(w - z)\, 
   \, \partial_{\bar z} \delta^2(0) = 0
$$
as $M_i \to \infty$.  

The first contraction in (\ref{tgraviton}) was calculated in section
 \ref{wardsection} and is equal to
$$
{\pi\alpha' k^2\over 4}
         \, \delta^2(w - z).
$$  
We calculate the second contraction in (\ref{tgraviton}) as follows, 
where we abbreviate $\bar k \equiv k_1 - i k_2$,
\begin{align*}
&{\pi \over 2}\,m^2 \,\expect{X^2(w)\,(\partial_z X)\, X(z)}
               + \mathit{PV} \\
&\quad = {\pi\over 2}\cdot m^2\cdot 2\cdot {i\over 2}
           \int {d^2p\over (2\pi)^2}\, e^{-ipx}
                   \int {d^2k\over (2\pi)^2} \, {\bar k
                              \over (k^2 + m^2)\, ((p - k)^2 + m^2)}
    + \mathit{PV} \\
&\quad =  {\pi\over 2}\, m^2\, i\, {1\over 4\pi}
              \int {d^2p\over (2\pi)^2}\, e^{-ipx}
              \int_0^1 dx \, {(1 - x)\bar p\over 
                             x(1-x)\, p^2 + m^2}
  + \mathit{PV} \\
&\quad =  {\pi\over 2}\, m^2\, i\, {1\over 4\pi}
              \int {d^2p\over (2\pi)^2}\, e^{-ipx}
              \int_0^{1/2} dx \, {\bar p\over 
                             x(1-x)\, p^2 + m^2} 
+ \mathit{PV}\\
&\quad =  {\pi\over 2}\, {i\over 4\pi}\, 2
     \int {d^2p\over (2\pi)^2}\, e^{-ipx}
           \int_{2\mu}^\infty {d\mu\over \mu^2}\, {m^2\over \sqrt{1 - 4m^2/\mu^2}}\,
               {\mu^2 \bar p\over p^2 + \mu^2} 
+ \mathit{PV}\\
&\quad =  {\pi\over 2}\, {i\over 4\pi}\, {2\over 8}\,
    \int {d^2p\over (2\pi)^2}\, e^{-ipx}
           \int_1^\infty {d\nu\over \nu^2}\,
       {1\over \sqrt{\nu^2 - 1}}\, 
        {4\, m^2\nu^2\bar p\over p^2 + 4 m^2 \nu^2} 
+ \mathit{PV}\\
&\quad \to  \left(\sum_i \eta_i^2\right){\pi\over 2}\, {i\over 4\pi}\, {2\over 8}
\int {d^2p\over (2\pi)^2}\, e^{-ipx}
\int_1^\infty {d\nu\over \nu^2}\,
       {1\over \sqrt{\nu^2 - 1}}\, \bar p \\
&\quad =  \left(\sum_i \eta_i^2\right){\pi\over 2}\, {i\over 4\pi}\, {2\over 8}
\int {d^2p\over (2\pi)^2}\, e^{-ipx}
\, \half\, B\left(\hhalf, 1\right)  \bar p \\
&\quad = (-1)\,{\pi\over 2}\, {i\over 4\pi}\, {2\over 8}
\int {d^2p\over (2\pi)^2}\, e^{-ipx}\, \bar p \\
&\quad = -{\pi\over 2}\, {i\over 4\pi}\, {2\over 8}\,
   2i \,\partial_z\delta^2(w - z) \\
&\quad = {1\over 16}\, \partial_z\delta^2(w - z).
\end{align*}
The arrow indicates the limit $m\to 0$ and $M_i\to \infty$, in which
the matter contribution vanishes an the Pauli-Villars contribution 
simplifies.  As usual, changes of variables performed in the above 
integrals are justified by their convergence when the Pauli-Villars
terms are included. 

Inserting this in (\ref{tgraviton}), we finally get
\begin{align*}
  \expect{T_{w\bar w} \tilde V_z \cdots} = {\pi\alpha' k^2\over 4}
         \, \delta^2(w - z)\, \expect {\tilde V_z\cdots}
     &+ {\pi \alpha' i k\over 8} \, \partial_z \delta^2(w - z)\,
        \expect{ c^z c^{\bar z}\, \partial_{\bar z} \tilde X\, e^{ik\tilde X}\cdots} \\
     & + {\pi \alpha' i k\over 8} \, \partial_{\bar z} \delta^2(w - z)\,
        \expect{ c^z c^{\bar z}\, \partial_{z} \tilde X\, e^{ik\tilde X}\cdots}.
\end{align*}
Inserting in the gauge invariance condition (\ref{graviton}), we find 
\begin{align*}
 0 =  {\pi\alpha' k^2\over 4}
         \, \delta\omega \expect {\tilde V\cdots} 
   &-{\pi \alpha' \, ie^{ij}k_j\over 8} \,(\partial_z\delta\omega) 
      \expect{ c^z c^{\bar z}\, \partial_{\bar z} \tilde X_i\, e^{ik\tilde X}\cdots} \\
  &-{\pi \alpha' \, ie^{ij}k_i\over 8} \,(\partial_{\bar z}\delta\omega) 
      \expect{ c^z c^{\bar z}\, \partial_{\bar z} \tilde X_j\, e^{ik\tilde X}\cdots}. 
\end{align*}
Since $\delta\omega$ is an arbitrary function, 
we find the following conditions on the momentum and the 
polarization.  
\begin{align*}
  0 &= k^2, \\
  0 &= e^{ij}k_j =  e^{ij}k_i.
\end{align*}
These are the usual mass shell and polarization conditions for the 
graviton vertex.

\section{Correspondence with Virasoro conditions}

The physical state condition (\ref{Vcondition}) 
\begin{align}
 0 = \expect{{\delta \tilde V\over \delta \hat g^{ij}}\,\hat g^{ij} \cdots}
   \delta \omega(\hat x)
   +{1\over 4\pi} \int d^2x\,\sqrt{\hat g} \expect { T_i^{\phantom{i}i}(x)\, \tilde V(\hat x) \cdots} \delta \omega(x) \label{Vcondition1}  
\end{align}
was derived by 
requiring that the functional integral be independent of the choice 
of gauge fixing function.  This condition was physically and 
mathematically 
well motivated from first principles 
in the functional integral approach, but 
does not obviously resemble the more familiar Virasoro conditions
on physical states.  
In particular, as we remarked, in  (\ref{Vcondition1}) 
only Weyl invariance led
to constraints on the physical insertions.
Diffeomorphism invariance was automatic and did not
contribute any constraints.  In the operator formalism, 
 the Virasoro constraints are usually 
motivated by a requirement of invariance under conformal 
transformations, 
which only correspond to a subclass of Weyl transformations
composed with specific diffeomorphisms.  It is therefore 
instructive to relate the two approaches.      

In this section
we will show that our constraint  (\ref{Vcondition1})
in fact implies the Virasoro
conditions. 

To see this, consider $\delta\omega$ of the special form
$$
  \delta\omega = - \half \,\nabla_i v^i,
$$
where $v^i$ is a vector field that is zero at $\hat x$,
conformal in a neighbourhood
$D$ of $\hat x$, and sufficiently 
well-behaved at infinity.  On the region $D$ where $v$ is conformal,  
we have, by definition of conformality,
$$
  \delta\omega \, g^{ij}= - \half \,\left(\nabla^i v^j + \nabla^j v^i\right)
     = \half \,h^{ij}.
$$
For this choice of $\delta\omega$, the  condition (\ref{Vcondition1})
becomes   
\begin{align*}
 0 &= \expect{\delta \tilde V\over \delta \hat g^{ij} \cdots}\,h^{ij}(\hat x)
  +{1\over 4\pi} \int_D d^2x\,\sqrt{\hat g} \expect { T_{ij}(x)\, \tilde V(\hat x)\cdots} h^{ij}(x) \\
&\quad
   +{1\over 4\pi} \int_{\bar D} d^2x\,\sqrt{\hat g} \expect {T_i^{\phantom{i}i}(x)\, \tilde V(\hat x)\cdots} 2\,\delta \omega(x) \\ 
&= \expect{\delta \tilde V\over \delta \hat g^{ij}}\,h^{ij}(\hat x)
  +{1\over 4\pi} \int_D d^2x\,\sqrt{\hat g} \expect { T_{ij}(x)\, \tilde V(\hat x)\cdots} h^{ij}(x),
\end{align*}
by virtue of the region of integration, 
since $T_i^{\phantom{i}i}(x)\, \tilde V(\hat x)$ can contribute at most 
contact terms.   Here $\bar D$ denotes the complement of $D$.  

But for the restricted region of integration, 
this is almost the Ward identity (\ref{vward}), remembering that $v^i$ is here
chosen zero at $\hat x$.  Subtracting (\ref{vward}) from this, the condition
becomes
\begin{align*}
  0 = {1\over 4\pi} \int_{\bar D} d^2x\,\sqrt{\hat g} \expect { T_{ij}(x)\, \tilde V(\hat x)\cdots} h^{ij}(x).
\end{align*}
In holomorphic coordinates, this gives
\begin{align*}
 0 = \int_{\bar D} d^2w\,
   \expect {\left(\partial_{\bar w}v^w\,T_{ww}
            + \partial_{w}v^{\bar w}\,T_{\bar w \bar w}\right) 
     \tilde V(z)\cdots}.
\end{align*}
Again, due to the region of integration, contact terms coming from  
$T_{w\bar w} \, \tilde V(z)$ do not contribute and have been dropped.  
A partial integration finally gives the physical state condition
\begin{align}
  0 = {1\over 2\pi i} \expect {\oint_{\partial D} \left(dw\,v^w\,T_{ww}
            - d\bar w\,v^{\bar w}\,T_{\bar w \bar w}\right) 
     c\bar c\,V(z)\cdots}.  \label {virasoro}
\end{align}
Given the basis
$$
  v^w \in \left\{(w- z)^{n+1}\,| \,n\ge 0\right\},
$$
of holomorphic vector fields on $D$, this is seen to
coincide with the usual Virasoro conditions.  We remind the reader that the 
contraction with $c\bar c$ reproduces the usual  
$a$-correction to $L_0$.    
  
The Virasoro conditions are preferable for actual calculations,
since they avoid the need for calculating complicated contact 
terms.  The above arguments can also be turned around to 
obtain certain contact terms from standard operator calculations, 
thus avoiding the rather complicated direct loop integral calculations of the 
previous section.   

We have shown that the condition (\ref{Vcondition1}) implies the Virasoro
conditions.  It might seem that the converse is not necessarily true, since
the $\delta \omega$ in (\ref{Vcondition1}) is arbitrary, whereas the class
of 
$\delta\omega$ needed
to derive the Virasoro conditions were harmonic on $D$, i.e., 
$\partial\bar\partial\,\delta\omega = 0$.  Comparing with the 
graviton condition in the previous section, we can see that the 
difference would matter 
for vertices that have mixed derivatives of $X$, the simplest of 
which is 
$(\partial\bar\partial X)\, e^{ikX}$, which would then give 
rise to terms proportional to $\partial\bar\partial\, \delta\omega$ 
in the physical state 
condition.  
However, these vertices vanish trivially by the equation of motion, and
are therefore of no interest.

\section{A covariant BRST approach}
\label{BRSTsection}

The BRST action for a generally covariant theory has subtleties
that are sometimes overlooked, and we will derive it carefully, 
obtaining an action that is different 
in important ways from the 
usual one.  In particular, the need for a fixed background metric 
$\bar g^{ij}$ in carrying out the Faddeev-Popov procedure will be
elucidated, and the independence of the resulting theory of 
this choice will be discussed.

For simplicity, we work on the plane, where the metric can be 
gauge-fixed completely and there are no remaining moduli.  
The aspects of the analysis that we would like to emphasize 
here do not involve the moduli.  

We apply the Faddeev-Popov and BRST procedures to the 
case at hand \cite{FP, BRST, Witten, Weinberg, nakahara}.  
Let $\bar g_{ij}$ be an arbitrary
background metric used for defining the measure as in section \ref{measuresection}.  The first 
ingredient we need is a functional integration measure that is invariant
under the gauge symmetries of the system.  In the current case, the 
symmetries are diffeomorphism invariance and conformal invariance.  
By our discussion in section \ref{measuresection}, both of 
these are preserved once we include Pauli-Villars partners for all 
fields.  The invariant measure is therefore
$$
  [dg]_{\bar g}\, [dB]_{\bar g} \,[dX]_{\bar g} \,
  [d\gamma]_{\bar g} \, [d\beta]_{\bar g} \, [d\chi]_{\bar g}. 
$$
As we discussed already, this measure is independent of the background
metric $\bar g$.  

Let
$\bar\phi^{kl}_n$
be an orthonormal basis of metric fields with respect to the
inner product 
$$
  \left<\bar\phi_n, \bar\phi_m \right>_{\bar g}
    = \int \sqrt{\bar g}\, \bar\phi_n^{ij}\, {\bar g}_{ik}
   {\bar g}_{jl} \, \bar\phi_m^{kl}.
$$
We note that $\tilde g^{ij}$ is different from the metric
 $\hat g_{ij}$ occurring in the gauge
fixing function $\delta (g^{ij} - {\hat g}^{ij})$.
Now consider the following set of constraint functions, which are
just the components of the metric field in the above basis.
$$
\bar\chi_n (g) \equiv \left<g , \bar\phi_n\right>_{\bar g}
 = \int \sqrt {\bar g} \,g^{ij}\, {\bar g}_{ik}
   {\bar g}_{jl} \, \bar\phi_n^{kl}.
$$
The Faddeev-Popov procedure instructs us to calculate the  
determinant of the matrix
$$
  \bar h_m(\bar\chi_n)
$$
where $\bar h_m$ is a fixed basis of generators of the symmetry group.
We may choose $\{\bar h_m\} = \{\bar v_p^i\} \cup \{ \bar\omega_q\}$, where
$\bar v_p^i$ is a $\bar g$-orthonormal basis of vector fields
generating diffeomorphisms, and $\bar\omega_q$ is
a $\bar g$-orthonormal basis of scalar fields generating Weyl 
rescalings.  

Note that the Faddeev-Popov procedure requires the 
basis of generators to be
fixed, independent of the variable of integration $g_{ij}$.  
We therefore could not have used $g_{ij}$ to define them.
Instead, an orthonormal basis with respect to 
the fixed background metric $\bar g_{ij}$ was used, but 
any fixed basis would have been acceptable.  A different choice would
lead at most to a different constant overall factor in the Haar measure 
on the group manifold.  This factor would be gauge slice independent
and would not affect physical expectation 
values.

Projecting the variation 
$$
  \delta g^{ij} = -\nabla_g^i v^j - \nabla_g^j v^i - 2 g^{ij} \delta\omega,
$$ 
onto $\bar \phi_n$, we find
$$
  \bar h_m(\bar\chi_n) = 
   \left\{\begin{array} {ll}
     - \left< \bar \phi^{ij}_n, \left(\nabla_g^k\bar v_m^l
      + \nabla_g^l\bar v_m^k\right) \right>_{\bar g}, & \qquad
         \bar h_m \equiv \bar v_p^i,\\
    -2 \left< \bar \phi^{ij}_n, g^{kl} \bar\omega_m \right>_{\bar g},
    &\qquad \bar h_m \equiv \bar\omega_q,
   \end{array}\right.
$$
The determinant of this matrix can be expressed in terms of a Grassmann
integral as
\begin{align}
 &  \int \prod d\bar c_m \prod d\bar b_n \,\exp{\sum \bar b_n \bar c_m\,  \bar h_m(\bar\chi_n)}
\nonumber \\
 &\qquad =
   \int [dc]_{\bar g} [dc^{\omega}]_{\bar g} [db]_{\bar g} 
      \,\exp\left\{-\sum \bar b_n \bar c_m\, 
            \left< \bar \phi^{ij}_n,\left( \nabla_g^k\bar v_m^l
         + \nabla_g^l\bar v_m^k\right) \right>_{\bar g}
        -2\sum \bar b_n \bar c^\omega_m\, \left< \bar \phi^{ij}_n, g^{kl} \bar\omega_m \right>_{\bar g}
     \right\} \nonumber\\
&\qquad =
   \int [dc]_{\bar g} [dc^{\omega}]_{\bar g} [db]_{\bar g} 
      \,\exp\left\{ 
            \int\sqrt{\bar g}\, b_{kl}\left(-\nabla_g^k c^l
         - \nabla_g^l c^k
        -2 \, g^{kl} c^{\omega} \right)
 \right\}, \label {FPgrassmann}
\end{align}
by definition of the measure for the ghost fields
discussed in section \ref{measuresection}, if we define $\bar b_n$ by
$$
\sum \bar b_n \,\bar\phi_n^{ij} \bar g_{ik}\bar g_{jl} = b_{kl},
$$ 
since $\bar\phi_n^{ij}\, \bar g_{ik}\bar g_{jl}$ is a $\bar g$-orthonormal
basis for the space of fields $b_{ij}$, 
and similarly for $c$ and $c^{\omega}$.

Diffeomorphism invariance, Weyl invariance and $\bar g$-independence
of the functional measure require the introduction of Pauli-Villars 
partners $v^i$, $\omega$ and $\rho_{ij}$ for the ghosts 
$c^i$, $c^\omega$ and $b_{ij}$.  Here, as before,
we use for example $v^i$ as a shorthand to denote
a set of massive fields whose total statistics is opposite to that of 
$c^i$, and whose bosonic elements have to be taken complex-valued 
to give a non-vanishing mass term.  For full details of the construction,
we refer back to section \ref{measuresection}.  
The full action for these fields will be stated below.  We remark 
that the bosonic $\rho_{ij}$, though complex-valued, are symmetric tensors,
not hermitian.   

We see that a correct application of the Faddeev-Popov procedure
gives the following action.
\begin{align*}
   S &\equiv \half \int \sqrt g\, g^{ij}\, \partial_i X\, \partial_j X
    + \int \sqrt {\bar g}\, b_{ab} \left(-\nabla^a_g c^b - \nabla^b_g c^a  
  -  2\, g^{ab}\, c^\omega\right)  \\
 &\qquad   - i\int \sqrt {\bar g}\,
B_{ij}\, (g^{ij} - \hat g^{ij}) + \cdots,
\end{align*}
where the dots indicate Pauli-Villars terms, to be fixed below from 
the requirement of BRST invariance.  
The action is of course coordinate-independent, but
it is important to note that it has 
an explicit 
background-dependence, as embodied in the presence of $\bar g$
in both the ghost and the gauge-fixing terms.
We shall see that no physical quantity will depend on $\bar g$.

This form of the action will be most convenient for discussing 
BRST-invariance.  However, for many calculations it will be 
more convenient to use a background-independent form of the 
ghost terms.  Such a representation will be derived in the next section, 
at the cost of obscuring BRST invariance.

The ghosts $c^i$ and $c^{\omega}$ may be identified with
the left-invariant one-forms on the gauge group, and gauge 
invariance of the original action can then be interpreted as invariance
under a BRST transformation $\delta_B$ which may be identified
with the exterior derivative on the gauge group \cite{nakahara}.  
As an exterior derivative in disguise, the BRST transformation
is then automatically nilpotent:
$$\delta_B^2 = 0.$$

The BRST transformations of the ghosts $c^i$ and $c^\omega$ 
are obtained from the 
Maurer-Cartan structure equations 
$$
  dc^m = -\half\,C^m_{np}\,c^nc^p
$$
that encode the exterior
derivatives of a basis $c^m$ of left-invariant one-forms
on the group, which depend on the commutation relations of the 
generators via the structure constants
$C^m_{np}$ \cite{nakahara}.  In this case the group is 
generated by vector fields and Weyl transformations, whose commutators
can be calculated relatively easily.  This gives the transformation
of $c^i$ and $c^\omega$.  The transformations of $b_{ij}$ and $B_{ij}$
are postulated in a standard way \cite{Witten, Weinberg}.   
The transformation of the other fields are obtained by simply lifting the
Lie algebra action onto the space of fields.  We obtain
\begin{align*}
  \delta_B X &= \mathcal{L}_{c} X = c^i\partial_i X, \\
 \delta_B g^{ij} &= \mathcal{L}_{c} g^{ij} 
 -   2\,g^{ij} c^\omega = -\nabla^i c^j - \nabla^j c^i -  2\,g^{ij} c^\omega, \\
\delta_B c^i &= - c^j\partial_j c^i = - \half\,\mathcal{L}_c c^i, \\
\delta_B c^{\omega} & = -c^i\partial_i c^\omega = -\mathcal{L}_{c} c^\omega, \\
  \delta_B b_{ij} &= -i B_{ij}, \\
  \delta_B B_{ij} &= 0.
\end{align*}
Here $\mathcal{L}_{c}$ denotes the Lie derivative, with respect to the 
vector field $c \equiv c^i\partial_i$, on the appropriate space of tensors.  
As is obvious from their re-expression as Lie derivatives, the 
ghost transformations are covariant.  But note that
the ghosts transform with different coefficients and opposite signs  
to the other fields.

Invariance of the action follows from the observation that the second and
third  terms can be written as
$$
  \int \delta_B \left(\sqrt{\bar g}\, b_{ab}\,g^{ab}\right) 
   + i\int \sqrt{\bar g}\,B_{ab}\,\hat g^{ab}
$$
The BRST variation of the first of these vanishes by 
$\delta_B^2 = 0$, which is satisfied by construction as noted above,
while the variation of the second vanishes trivially.  

The Pauli-Villars fields are all taken to transform as tensors
\begin{align*}
\delta_B \chi &= \mathcal{L}_{c} \chi = c^i\partial_i \chi, \\
 \delta_B \gamma^{ij} &= \mathcal{L}_{c} \gamma^{ij} 
 -   2\,\gamma^{ij} c^\omega,  \\
\delta_B \beta_{ij} &= \mathcal{L}_{c} \beta_{ij}, \\ 
\delta_B v^i &= \mathcal{L}_{c} v^i = [c, v] \\
\delta_B \omega &= \mathcal{L}_{c} \omega = c^i\partial_i \omega, \\
\delta_B \rho_{ij} &= \mathcal{L}_c \rho_{ij}.
\end{align*}
Note that $B$ and $\beta$ do not transform in the same way, and neither
do the partners $c$ and $v$, nor $c^\omega$ and $\omega$.  Also, note
that $\gamma$ transforms under conformal transformations in the same
way as $g$.    

The Pauli-Villars regularization will require small masses $m$ to be 
given to the ghost fields, and large masses to their auxiliary Pauli-Villars
fields.  The form of the mass terms are taken to be
\begin{align*}
 m \int \sqrt g \, \left(\sqrt g\, dx^1 \wedge dx^2 \right) (c, c)
   &=  m \int \sqrt g\left(\sqrt g\,\epsilon_{kl}\right) c^k c^l 
\end{align*}
and 
\begin{align*}
 m \int \sqrt g \, \left(\sqrt g\,\epsilon_{kl}\right) g^{kp}g^{lq}g^{ij}
   \, b_{ki}b_{lj}
\end{align*}
respectively.  These are diffeomorphism-invariant due to 
coordinate-invariance of the density 
$\sqrt g\, dx^1 \wedge dx^2$.
The full action is then 
\begin{align}
   S &\equiv \half \int \sqrt g\, g^{ij}\, \partial_i X\, \partial_j X
    + \int \sqrt {\bar g}\, b_{ab} \left(-\nabla^a_g c^b - \nabla^b_g c^a  
  -  2\, g^{ab}\, c^\omega\right)  \nonumber\\
 &\quad   - i\int \sqrt {\bar g}\,
B_{ij}\, (g^{ij} - \hat g^{ij})  \nonumber\\
&\quad +  \half \int \sqrt g\left( g^{ij}\, \partial_i \bar \chi\, \partial_j \chi + M_\chi^2 \,\bar\chi \chi\right)
+ \half\int \sqrt {g}\left\{ \bar\rho_{ab} \left(-\nabla^a_g v^b - \nabla^b_g v^a  
  -  2\, g^{ab}\, \omega\right) + \mathrm{c.c.}\right\} \nonumber\\
 &\quad + M_v  \int \sqrt g\left(\sqrt g\,\epsilon_{kl}\right)\bar v^k v^l 
+ 
 M_b \int \sqrt g \, \left(\sqrt g\,\epsilon_{kl}\right) g^{kp}g^{lq}g^{ij}
   \, \bar \rho_{ki}\rho_{lj} \nonumber \\
&\quad   - {i\over 2}\int \sqrt {g}\left(
\bar \beta_{ij}\,\gamma^{ij} + \mathrm{c.c.}\right).  \label{fullBRST}
\end{align}
It is important to note that the 
Pauli-Villars terms are taken to be diffeomorphism invariant, independent
of $\bar g$, in contrast with the original terms.  As a result, they are
invariant under the component of the BRST transformation
parameterized by the vector $c^i$, since this simply acts as an
infinitesimal diffeomorphism on the Pauli-Villars terms.  Notice,
though, that the mass terms break the Weyl component of the 
BRST transformation generated by $c^\omega$.  This will be 
origin of the conformal anomaly in this approach.  

In the above action, we have assumed the shorthand of
writing only one Pauli-Villars partner for each ordinary field.  In 
practice we usually need a few, enough to satisfy the required
conditions on the statistics and the masses \cite{myself}.  

An important property of the current construction is that the 
measure is BRST-invariant.  This is due to certain perhaps unexpected
cancellations of the BRST variations of different factors.  First,
note that trivially
$$
  \delta_B [dB] = 0.
$$
Also
$$
  \delta_B \left([db]\,[dB]\right) = 0,
$$
by the elementary property $db_n\wedge db_n = 0$ of the exterior
product.  
Also, per the discussion of section \ref{measuresection}, we have
$$
  \delta_B \left([dX]\,[d\chi]\right) = \delta_B \left([dg]\,[d\gamma]\right)
   = \delta_B \left([dc^\omega]\, [d\omega]\right) = 0,
$$
since in each case the two factors transform with opposite signs due to 
opposite statistics.  The same is true for the following pair of factors,
even though they are not Pauli-Villars partners:
$$
  \delta_B\left([\rho]\,[\beta]\right) = 0.
$$
Finally, a slightly nontrivial calculation shows that
$$
  \delta_B \left([dc]\,[dv]\right) = 0.
$$
Even though the BRST transformations of $c^i$ and $v^i$ differ
by a factor of $2$, this is compensated by the fact that 
$\delta_B c^i$ is quadratic in $c^i$ while $\delta_B v^i$ is
linear.  

We now consider insertions.  Before gauge fixing, each
insertion contributes
$$
  \int d^2x\,\sqrt g\,V(x)
$$
to the path integral.  This is diffeomorphism-invariant, though not 
in general Weyl-invariant; requiring the absence of the 
corresponding quantum anomaly will 
determine the physical state condition on $V$.  After using diffeomorphism
invariance of the full path integral to fix the metric to a fixed 
$\hat g$ up to a Weyl transformation, a finite set of diffeomorphisms,
Weyl related to global conformal transformations, may remain that can be
used to fix the positions $\hat x$ 
of a finite set of insertions.  For each of these,
the gauge fixing function may be chosen as
$$
  \delta (x - \hat x)
$$
and the Faddeev-Popov determinant contribution may be written as
\cite{polchinski}
$$
  \int d^2 \eta\, e^{\eta_i c^i(x)},
$$
where $\eta_i$ is an anti-commuting covector.  A simple calculation 
then gives the following contribution for a fixed insertion
\begin{align}
   \int d^2x\,\sqrt g\,V(x)\,\delta (x - \hat x) \int d^2 \eta\, 
     e^{\eta_i c^i(x)} = \sqrt {g(\hat x)}\, c^1(\hat x)\,c^2(\hat x)\,
     V(\hat x). \label {insertion}
\end{align}
As an aside, it is worth noting that this can be rewritten 
in terms of a measure that 
is both diffeomorphism- and Weyl invariant as follows:
$$
  \int d^2x \sqrt g\,V(x) \, \delta_g(x - \hat x)
   \int {d^2\eta_i\over \sqrt g}\, e^{\eta^i g_{ij}\, c^j(x)}, 
$$
which shifts all the Weyl variance to the integrand.

\section{Covariantizing the ghost term}

As discussed in the previous section,
the full action (\ref{fullBRST}) is BRST invariant except for
the Pauli-Villars mass terms that will contribute the conformal anomaly.  
However, notice that the 
ghost kinetic term depends on the background metric $\bar g$.
For some calculations, it is more convenient to replace this
term with a background-independent form.  In this section we 
will show that one can do this while making a corresponding 
change in the $\beta$-$\gamma$ term, at the cost of obscuring 
the manifest BRST invariance of the action.  

We first consider the partition function without insertions. 
Let $\bar\phi_m$ be an orthonormal basis for the space of 2-tensors with 
respect to the $\bar g$-inner product and $\phi_m$ a basis with respect 
to the $g$-inner product.  Similarly, let $\bar\varphi_n$ and $\varphi_n$ 
be orthonormal bases
for the space $V^1\oplus V^0$, where $v^1$ denotes the space of vector
and $v^0$ the space of scalar fields, with respect to the $\bar g$-inner
product and the $g$-inner product respectively.  Also, denote 
$$
  i_g (v \oplus \omega)\equiv \nabla_g^i v^j + \nabla_g^j v^i
   - 2\, g^{ij}\omega.
$$
The relevant factors in (\ref{fullBRST}) can be represented as
\begin{align*}
 & \int [db]_{\bar g}\,[dc]_{\bar g}\,[d\rho]_{\bar g}\,[dv]_{\bar g}
    \,[dg]_{\bar g}\,[d\gamma]_{\bar g}\,[dB]_{\bar g}\,[d\beta]_{\bar g}\,
     e^{ \braket{b}{i_{g} c}_{\bar g}} e^{\half\left(\braket {\beta}{\gamma}_g + \mathrm{c.c.}\right)}\cdots \\
&\quad =  \int [db]_{\bar g}\,[dc]_{\bar g}\,[d\rho]_{\bar g}\,[dv]_{\bar g}
    \,[dg]_{g}\,[d\gamma]_{g}\,[dB]_{g}\,[d\beta]_{g}\,
     e^{ \braket{b}{i_{g} c}_{\bar g}} e^{\braket {\beta}{\gamma}_g}\cdots \\
&\quad =  \int [db]_{\bar g}\,[dc]_{\bar g}\,[d\rho]_{\bar g}\,[dv]_{\bar g}
    \,[dg]_{g}\,[dB]_{g}\,
     e^{ \braket{b}{i_{g} c}_{\bar g}} \cdots \\
&\quad =  \int [d\rho]_{\bar g}\,[dv]_{\bar g}
    \,[dg]_{g}\,[dB]_{g}\,
     \det \braket{\bar \phi_m}{i_g \bar\varphi_n}_{\bar g} \cdots \\
&\quad =  \int [d\rho]_{\bar g}\,[dv]_{\bar g}
    \,[dg]_{g}\,[dB]_{g}\,
     \det \braket{\bar\phi_m}{\phi_n}_{\bar g}\,
     \det \braket{\phi_n}{i_g \varphi_p}_{g}\,
      \det \braket{\varphi_p}{\bar\varphi_q}_{g}\cdots \\
&\quad =  \int [d\rho]_{\bar g}\,[dv]_{g}
    \,[dg]_{g}\,[dB]_{g}\,
     \det \braket{\bar\phi_m}{\phi_n}_{\bar g}\,
     \det \braket{\phi_n}{i_g \varphi_p}_{g}\,
      \cdots \\
&\quad =  \int [d\rho]_{g}\,[dv]_{g}
    \,[dg]_{g}\,[dB]_{g}\,
    \det \braket{\phi_p}{\bar\phi_m}_{\bar g}\,
     \det \braket{\bar\phi_m}{\phi_n}_{\bar g}\,
     \det \braket{\phi_n}{i_g \varphi_p}_{g}\,
      \cdots \\
&\quad =  \int [d\rho]_{g}\,[dv]_{g}
    \,[dg]_{g}\,[dB]_{g}\,
    \det \braket{\phi_p}{\phi_n}_{\bar g}\,
     \det \braket{\phi_n}{i_g \varphi_p}_{g}\,
      \cdots \\
 &\quad = \int [db]_{g}\,[dc]_{g}\,[d\rho]_{g}\,[dv]_{g}
    \,[dg]_{g}\,[d\gamma]_{g}\,[dB]_{g}\,[d\beta]_{g}\,
      e^{\half\left(\braket {\beta}{\gamma}_{\bar g} + \mathrm{c.c.}\right)} e^{ \braket{b}{i_{g} c}_{g}} 
     \cdots \\
 &\quad = \int [db]_{\bar g}\,[dc]_{\bar g}\,[d\rho]_{\bar g}\,[dv]_{\bar g}
    \,[dg]_{\bar g}\,[d\gamma]_{\bar g}\,[dB]_{\bar g}\,[d\beta]_{\bar g}\,
      e^{\half\left(\braket {\beta}{\gamma}_{\bar g} + \mathrm{c.c.}\right)} e^{ \braket{b}{i_{g} c}_{g}}
     \cdots 
\end{align*} 
where we have treated various infinite-dimensional determinants 
informally.  
In the second line we used $\bar g$-invariance of the combined
matter-Pauli-Villars measures.  In the third line we integrated over 
$\beta$ and $\gamma$ to obtain $1$.  In the sixth line we used 
$$
[dv]_{g} = [dv]_{\bar g}\,\det \braket{\varphi_p}{\bar\varphi_q}_{g},
$$
and then 
$$
[d\rho]_{\bar g} = [d\rho]_{g}\,\det \braket{\phi_p}{\bar\phi_m}_{\bar g}.
$$
We obtain the following action:
\begin{align}
   S &\equiv \half \int \sqrt g\, g^{ij}\, \partial_i X\, \partial_j X
    + \int \sqrt {g}\, b_{ab} \left(-\nabla^a_g c^b - \nabla^b_g c^a  
  -  2\, g^{ab}\, c^\omega\right)  \nonumber\\
 &\quad   - i\int \sqrt {\bar g}\,
B_{ij}\, (g^{ij} - \hat g^{ij})  \nonumber\\
&\quad +  \half \int \sqrt g\left( g^{ij}\, \partial_i \bar \chi\, \partial_j \chi + M_\chi^2 \,\bar\chi \chi\right)
+ \half\int \sqrt {g}\left\{ \bar\rho_{ab} \left(-\nabla^a_g v^b - \nabla^b_g v^a  
  -  2\, g^{ab}\, \omega\right) + \mathrm{c.c.}\right\} \nonumber\\
 &\quad + M_v  \int \sqrt g\left(\sqrt g\,\epsilon_{kl}\right)\bar v^k v^l 
+ 
 M_b \int \sqrt g \, \left(\sqrt g\,\epsilon_{kl}\right) g^{kp}g^{lq}g^{ij}
   \, \bar \rho_{ki}\rho_{lj} \nonumber \\
&\quad   - {i\over 2}\int \sqrt {\bar g}\left(
\bar \beta_{ij}\,\gamma^{ij} + \mathrm{c.c.}\right).  \label{covBRST}
\end{align}
Compared with the original action (\ref{fullBRST}), the 
$b$-$c$ term is now $\bar g$-independent, while the $\beta$-$\gamma$ term
now depends on $\bar g$.

We now consider the insertions.   These are, in fact, unaffected 
by the above manipulations, since
\begin{align*}
   \int [dc]_{\bar g}\,[dv]_{\bar g}\int d^2 \eta\, 
     e^{\eta_i c^i(x)}
  &= \int [dc]_{g}\,[dv]_{g}\int d^2 \eta\, 
     e^{\eta_i c^i(x)}
\end{align*}
or 
\begin{align*}
   \int [dc]_{\bar g}\,[dv]_{\bar g}\int d^2 \eta\, 
     e^{\eta_i \sum \bar\varphi_n^i \bar c_n}
  &= \int [dc]_{g}\,[dv]_{g}\int d^2 \eta\, 
     e^{\eta_i \sum \varphi_n^i c_n}
\end{align*}
The insertions are therefore still given by (\ref{insertion}).

\section{BRST anomalies}

It is 
straightforward to check, from the BRST-invariant form 
(\ref{fullBRST}) of the action, that
\begin{align}
 \delta_B S
   = - {1\over 4\pi}\int
    d^2 x\, \sqrt g \,(- 2\,c^\omega)\, T^i_{\phantom{i}i}(\chi, \rho, v, \omega).  \label{deltabS}
\end{align}
Only the contributions to $T^i_{\phantom{i}i}$ from the
Pauli-Villars mass terms contribute here, since 
the $c^\omega$-variance of the ghost term cancels
that of the $(B, g)$ gauge fixing term.  Also note
that when using the form
(\ref{fullBRST}), the $(\beta, \gamma)$ term is conformally
invariant and therefore does not contribute a $c^\omega$ 
variance to $\delta_B S$.

We may obtain a current associated with $\delta_B$ by 
a standard procedure as follows.  Write the change of variables 
formula for the path integral
$$
  \int [d\phi]' \, e^{- \tilde S(\phi')} 
   = \int [d\phi] \, e^{- \tilde S(\phi)} 
$$  
where $\phi$ stands for all the fields in the theory, and
 where we take the change of variables 
$$
  \phi'(x) = \phi(x) + \epsilon(x)\,\delta_B\phi(x),
$$
where $\epsilon(x)$ denotes a Grassmann-valued function.  
Since this is a symmetry of the 
massless action for $\epsilon$
constant, the variation of this part of the action is
proportional
to derivatives of $\epsilon$ which, after a partial
integration, would give rise to a conserved current \cite{polchinski, myself}
in the absence of additional mass terms.  
Here, the presence of the Pauli-Villars mass terms will 
give additional contributions.

We note that, as discussed already, 
the measure is invariant under $\delta_B$.
With this in mind, only the variation of the action contributes
in the above formula.  We start by taking the non-covariantized version
(\ref{fullBRST}) of the action, and we obtain 
\begin{align}
  0 &=  {1\over 4\pi}\int d^2 x\,\biggl\{(2\,\nabla_g^i\epsilon)
   \, \expect{\left( \sqrt g\,  c^j \,
     T_{ij}(X, \chi, \rho, v, \omega) + \sqrt {\bar g}\,b_{ik}\,c^j\partial_j c^k\right) \cdots
      }  \nonumber\\
  &\qquad\qquad \qquad\qquad + \epsilon\expect{ \sqrt g \left(- 2 \,c^\omega \right) T^i_{\phantom{i}i}(\chi, \rho, v, \omega)\cdots   } \biggr\}
\label{conserv}
\end{align}
where the dots denote possible additional insertions outside the 
support of $\epsilon(x)$.  
To derive this formula, note that for the fields $(X, \chi, \rho, v, \beta, \gamma, \omega)$, 
the transformation under the $c^i$-indexed part of 
$\delta_B$ is identical to an infinitesimal
diffeomorphism parameterized by $c^i$, and we obtain the energy-momentum
tensor $T_{ij}(X, \chi, \rho, v)$ in the usual way.  Note also
that the $(\beta, \gamma)$ contribution is zero, since this term is both 
diffeomorphism and Weyl invariant
and has no derivatives, so that no $\epsilon$-derivatives are contributed.
Next, note that terms proportional to $\epsilon(x)$ cancel between the 
$(b, c)$ and $(B, g)$ terms as in the case of constant $\epsilon$, 
and only the terms containing derivatives of
$\epsilon$ in the variation of the ghost terms remain.  

Note that, by the construction in section \ref{BRSTsection}, changing $\bar g^{ij}$ at 
worst changes the overall normalization of the Haar measure.  
As a result, the path integral is
independent, up to a possible overall gauge slice-independent factor,
of the background metric $\bar g^{ij}$, as long as 
any further ghost insertions are 
in the specific format (\ref{insertion}) required for fixed 
vertex operators.

However, the above expression contains an explicit instance
of $\bar g$ that is not in an overall normalization position, 
in seeming contradiction with this statement.  
This is resolved by noting
that from the $(b, c)$ term in the action it is easy to see that 
propagators involving the combination $\sqrt {\bar g}\, b_{ij}$ are 
independent of $\bar g$.   

This remark also explains why the 
above expression will be finite due to cancellation
of self-contraction divergences between fields and their Pauli-Villars partners, as expected for our covariant regularization.  Indeed
even though the $(\rho, v)$ term has no $\bar g$-dependence, the
combination $\sqrt{ \hat g}\, \rho_{ij}$ will have the same propagator as $\sqrt {\bar g}\, b_{ij}$.

Note that we are free to choose $\bar g^{ij} = \hat g^{ij}$ for a fixed 
choice $\hat g^{ij}$ of gauge slice. 
However, slice independence of the path integral is only valid
as long as we vary $\hat g$ independently of $\bar g$.  
Otherwise the Haar normalization
factor would become dependent on $\hat g$, and gauge slice independence of 
the path integral would be broken.  

To avoid this issue, and to avoid dealing with two different metrics
in the resulting conservation law, 
it is worth comparing the above identity with the one we would have
obtained if we had started from the covariantized action (\ref{covBRST})
instead.  It would be 
\begin{align*}
  0 &=  {1\over 4\pi}\int d^2 x\,\biggl\{(2\,\nabla_g^i\epsilon)
   \, \expect{\left( \sqrt g\,  c^j \,
     T_{ij}(X, \chi, \rho, v, \omega) + \sqrt {g}\,b_{ik}\,c^j\partial_j c^k\right) \cdots
      }  \\
  &\qquad\qquad+ \epsilon\biggl< \left(- 2 \,c^\omega \right) \biggl( \sqrt g\, T^i_{\phantom{i}i}(\chi, \rho, v, \omega)  \\
  &\qquad\qquad \qquad\qquad
 + \sqrt {g}\, b_{ab} \left(-\nabla^a_g c^b - \nabla^b_g c^a  
  -  2\, g^{ab}\, c^\omega\right)
 + {i\over 2} \,\sqrt {\bar g}\left(
\bar \beta_{ij}\,\gamma^{ij} + \mathrm{c.c.}\right)
\biggr) \cdots
   \biggr> \biggr\},
\end{align*}
and since the actions were argued to be equivalent, this identity should
be equivalent to the first identity (\ref{conserv}) above
as long as 
any further ghost insertions are 
in the specific format (\ref{insertion}).  

To see that this is true, note that
the ghost term in the first line now depends on $\sqrt g$, but now so 
does the corresponding term in the action (\ref{covBRST}), 
so that propagators
involving $\sqrt g\, b_{ij}$ will be the same as propagators involving 
$\sqrt {\bar g}\, b_{ij}$ in the first version above.  
The new terms on the last line will compensate each other.  
In particular,
from the action we can read off that 
the contraction of $\sqrt g\,b_{ab}$ with  $-\nabla^a_g c^b - \nabla^b_g c^a  
  -  2\, g^{ab}\, c^\omega$ is the same as the contraction of 
$ -{i\over 2}\,\beta_{ij}$ with $\gamma^{ij}$.  As a result,  
self-contractions
on the last line cancel.  Furthermore, if $\cdots$ contains vertices with accompanying
$c$-ghost factors, these may contribute contractions with $b_{ab}$ in the first 
term, but the result is then proportional to 
$-\nabla^a_g c^b - \nabla^b_g c^a  
  -  2\, g^{ab}\, c^\omega$, which vanishes by the equations of motion 
in the expectation value under the stated assumption on the content
of $\cdots$.   We also assume that $\cdots$ contains no $(\beta, \gamma)$
dependent insertions.  As a result of all these arguments, both terms
on the last line may be dropped completely, and we obtain 
a nice identity without explicit $\bar g$-dependence.  
\begin{align*}
  0 &=  {1\over 4\pi}\int d^2 x\,\biggl\{(2\,\nabla^i\epsilon)
   \, \expect{\left( \sqrt g\,  c^j \,
     T_{ij}(X, \chi, \rho, v, \omega) + \sqrt {g}\,b_{ik}\,c^j\partial_j c^k\right) \cdots
      }  \\
  &\qquad\qquad\qquad\qquad + \epsilon\expect{ \left(- 2 \,c^\omega \right)\, \sqrt g\, T^i_{\phantom{i}i}(\chi, \rho, v, \omega)\,  
\cdots
   } \biggr\},
\end{align*}  
Performing the integral over $g$, which applies the gauge-fixing delta
function fixing $g = \hat g$, we get
\begin{align*}
0 =  \int d^2 x\, \sqrt {\hat g}\, \epsilon(x)\,
   \, \expect {\left\{-\nabla^i \left(c^j 
      T_{ij}(X, \chi, \rho, v, \omega) + b_{ik}\,c^j\partial_j c^k
      \right)  - c^\omega  T^i_{\phantom{i}i} (\chi, \rho, v, \omega)\right\}\cdots}_{\hat g}.
\end{align*}
Since $\epsilon(x)$ is arbitrary, we find the covariant conservation law
$$
  \expect{\nabla^i j_i \cdots}_{\hat g} = -\expect {c^{\omega} \,  T^i_{\phantom{i}i}(X, \chi, \rho, v, \omega)
    \cdots}_{\hat g}
$$
as long as no other insertion is at the point
in question.  
Here the BRST current is 
$$
  j_i \equiv c^j 
      T_{ij}(X, \chi, \rho, v, \omega) + b_{ik}\,c^j\partial_j c^k
$$
Remembering that
$c^j\partial_j c^k = \half\,\mathcal{L}_c c^k$, we note that 
$j_i\,dx^i$ is indeed a coordinate-invariant, true tensor quantity. 
Also note that the anomalous quantity 
$T^i_{\phantom{i}i} (\chi, \rho, v, \omega)$ on the right hand side
does not include contributions from $(B, g, \beta, \gamma)$, even though
the corresponding terms in (\ref{covBRST}) are not Weyl invariant.

When no other insertion coincides with $ T^i_{\phantom{i}i}$, 
we may replace \cite{myself}
$$
  T^i_{\phantom{i}i} \to -{c\over 12}\, R,
$$
where $R$ is the curvature of $\hat g$ and $c$ is the total central charge, 
and we get
$$
  \expect{\nabla^i j_i \cdots} = {c\over 12}\,R \expect {c^{\omega} \cdots}.
$$
Finally, $c$ will be calculated in section \ref{anomalysection} and shown to be zero
in dimension $d = 26$, in which
case we indeed get a non-anomalous covariant conservation law.  
  
We may also use the equation of motion, valid in expectation values,
to replace 
$$
  c^\omega \to -\half \left(\nabla^i c^j + \nabla^j c^i\right) 
$$
on the right hand side.

It is essential to understand that, just like 
the energy-momentum tensor in the covariant regularization
\cite{myself}, this current
is \textit{finite}.  Divergences arising from self-contractions
of the second term are exactly canceled by self-contractions
of the Pauli-Villars contribution $c^j  T_{ij}(\rho, v)$.  As a result, no 
covariance-spoiling renormalization is needed, in contrast 
with the usual operator formalism \cite{polchinski},.

Since the Pauli-Villars regularization is coordinate-independent,
and provides finite insertions without the need for further 
renormalizations \cite{myself}, the BRST current $j_i\, dx^i$ is
by construction a true coordinate-independent tensor quantity.  
In this the current formalism differs from the operator formalism,
 where 
the renormalizations depend on a coordinate choice, leading to
an anomalous transformation law for the current \cite{polchinski}.  The $j$
constructed here does not transform anomalously but instead 
contains extra contributions.  To see this clearly, take
 $\hat g$ to be the 
flat metric, so that we may write
\begin{align}
  j_z &= 
 c^z
      T_{zz}(X, \chi, \rho, v, \omega) + b_{zz}\,c^z\partial_z c^z \\
 &\quad
      + c^{\bar z}T_{z\bar z}(\chi, \rho, v, \omega) + b_{z\bar z}\,c^z\partial_z c^{\bar z}  
+ b_{z\bar z}\,c^{\bar z}\partial_{\bar z} c^{\bar z} + b_{zz}\,c^{\bar z}\partial_{\bar z} c^{z}.  \label{flatj}
\end{align}
The terms on the second line do not vanish in our covariant 
regularization, since they may contribute contact terms to expectation 
values.  It is obvious
that dropping some of these terms, as is done in the usual operator formalism,
would lead to a non-covariant, 
non-tensor object.

We may follow a similar procedure to obtain
Ward identities involving insertions, where the
current acts as generator for the  $\delta_B$.  Concretely,
if $\mathcal{O}(\hat x)$ is an isolated insertion, the same 
argument, using an $\epsilon(x)$ that is nonzero only in a neighbourhood
of $\hat x$ that contains no other insertions, gives
the Ward identity
\begin{align}
0 = \epsilon(\hat x)\expect{\delta_B \mathcal{O}\cdots}_{\hat g} -
  {1\over 2\pi} \int d^2 x\, \sqrt {\hat g}\, \epsilon(x)\,
   \, \expect {\mathcal{O}(\hat x)\left\{\nabla^i j_i  + c^\omega  T^i_{\phantom{i}i}(\chi, \rho, v, \omega)\right\} \cdots}_{\hat g}  \label{classicalward}
\end{align}
If we take $\epsilon(x) = \epsilon$ on a neighbourhood of $\hat x$ and 
$\epsilon \to 0$ outside this neighbourhood,
the first term is then a total derivative, which 
can be integrated to give a BRST charge.  
As noted above in (\ref{flatj}), 
this charge will contain terms proportional to boundary
integrals of 
$$ T_{z\bar z}(X, \chi, \rho, v, \omega) + b_{z\bar z}\,c^z\partial_z c^{\bar z}  
+ b_{z\bar z}\,c^{\bar z}\partial_{\bar z} c^{\bar z} + b_{zz}\,c^{\bar z}\partial_{\bar z} c^{z}$$
that are not there in the 
usual operator formalism.

However, since these extra 
 contributions
 may only contribute contact terms, and since the 
insertion in (\ref{classicalward}) 
 does not touch the boundary, we may drop them to obtain the usual
holomorphic BRST charges.  
Note that the Pauli-Villars contributions in
$$
   c^z T_{zz}(X, \chi, \rho, v, \omega)
$$
cannot be dropped, even though they can only give rise to contact 
terms when contracted with the insertion, since they also
contribute self-contractions that cancel divergences arising from 
self-contractions of the matter and the ghost terms \cite{myself}.

In (\ref{classicalward}), it is important to note that the second term 
on the right hand side is not a total derivative, and cannot be 
directly converted into a surface integral.  Since it can touch the 
insertion, we cannot simplify it as we did in the case 
of current conservation above.   
Due to the presence of this term, the 
 formula (\ref{classicalward}) generates the 
\textit{classical}, non-anomalous BRST transformation $\delta_B$.  

This is indeed the expected behaviour of a coordinate-invariant, 
Pauli-Villars
based regularization, as we have seen also in the case of the energy-momentum
tensor Ward identities in section \ref{wardsection}.  As there, this
Ward identity can be related to the usual operator formalism identity
\cite{polchinski} by non-covariant operator redefinitions, which would
be as unnatural in the present context as  they were in section 
\ref{wardsection}.  
Possible anomalies, leading to
 the physical
state conditions, present themselves
in a different way, as we shall see below.

At first glance it may seem as if the above Ward identity cannot 
possibly be correct.  Since $j_i$ contains no instances of the 
field $B_{ij}$ conjugate  to $g^{ij}$, it would appear that the formula
(\ref{classicalward}) cannot be used to generate transformations of
$\mathcal O = g^{ij}$.  However, 
one should remember that we are deriving identities that 
are true inside expectation values.  
For example, taking $\mathcal{O} = g^{ij}$, the right hand side of 
(\ref{classicalward}) vanishes, while the left hand side is just
the expectation value of $\delta_B g^{ij}$, so that 
we obtain
\begin{align}
  \expect{-\nabla^i c^j - \nabla^j c^i - 2\, c^\omega \hat g^{ij}
   \cdots}_{\hat g} = 0,  \label{cward}
\end{align}
which simply states that the expectation values satisfy the 
classical equations of motion.

The physical state condition may be recast as follows in the 
BRST language. 
We restate the requirement,
considered in section \ref{physicalsection},
that $$\expect {V({\hat x}) \cdots}$$ be independent of the choice of
gauge fixing function indexed by $\hat g$.
\begin{align*}
   0 &=
    -i \int d^2 x\,\sqrt{\bar g}\expect{\tilde V(\hat x)\,B_{ij}(x) \cdots}\delta\hat g^{ij}(x)\\
   &= \int d^2 x\,\sqrt{\bar g}\expect {\tilde V(\hat x)\,\delta_B b_{ij}(x) \cdots}\delta\hat g^{ij}(x) \\
   &= \int d^2 x\,\sqrt{\bar g}\int d\mu\, \tilde V(\hat x)\,e^{-S}\,(\delta_B b_{ij}(x))\,\delta\hat g^{ij}(x)\cdots \\
   &= -\int d\mu \left((\delta_B \tilde V)(\hat x)\, e^{-S}-
     \int d\mu \, V(\hat x)\, (\delta_B  S)\,e^{-S}\right)\int d^2x\,\sqrt{\bar g}\, b_{ij}(x)\,\delta\hat g^{ij}(x) \cdots + \cdots\\
  &= -\expect {\biggl(\delta_B - \delta_B S\biggr) \tilde V(\hat x) \int d^2x\,\sqrt{\bar g}\,b_{ij}(x)\,\delta\hat g^{ij}(x) \cdots} + \cdots,
\end{align*}
where we may use either action (\ref{fullBRST}) or (\ref{covBRST}).
Again we have used that the measure is BRST-invariant, or
$$
  \int d\mu \, \delta_B (\cdots) = 0.
$$
The fact that 
we shall obtain the correct physical state conditions
constitutes an independent
check that this is indeed 
correct.

It is now obvious where the quantum condition differs from the 
classical one.  Indeed, as noted in 
(\ref{deltabS}), the Pauli-Villars regularization contributes
mass terms to $S$, so that 
\begin{align*}
  \delta_B  S 
   &= - {1\over 4\pi}\int
    d^2 x\, \sqrt g \,(- 2\,c^\omega)\, T^i_{\phantom{i}i}(\chi, \rho, v, \omega)
    \ne 0.
\end{align*}
The trace of the energy-momentum tensor
contributes contact terms in the limit of infinite Pauli-Villars
masses.  
For example, for the tachyon
$$\tilde V = \sqrt g \, \epsilon_{ij}\, c^i c^j\, e^{ik\tilde X},$$
we obtain, in conformal coordinates,
$$
  \delta_B \,(\sqrt g \, c\bar c) = 2\, c^\omega\,(\sqrt g\, c\bar c),
$$
where terms containing $\partial c$ and $\bar \partial \bar c$ have 
canceled between $\delta_B\sqrt g$ and $\delta_B (c\bar c)$, so that
\begin{align*}
  0 = \expect{\left(2\, c^\omega(z)\, \tilde V(z) + {1\over 4\pi}\int
    d^2 w\, (-4\, c^\omega(w))\left(T_{w\bar w} + T_{\bar w w}\right) \tilde V(z) 
     \right) \int d^2u\,b_{ij}(u)\,\delta\hat g^{ij}(u) \cdots}.
\end{align*}
Choosing $\delta\hat g^{ij} = -2\,\hat g^{ij}\,\delta\omega$, 
and then contracting 
$$
  c^\omega(z) \,b_i^{\phantom{i}i}(u) \sim \delta^2 (z, u)
$$ 
gives 
\begin{align*}
  0 = \expect{\left(2\, \delta\omega(z)\, \tilde V(z) + {1\over 4\pi}\int
    d^2 w\, (-4\, \delta\omega(w))\left(T_{w\bar w} + T_{\bar w w}\right) \tilde V(z) 
     \right) \cdots},
\end{align*}
which is exactly the same condition as the one obtained for the tachyon
in 
section \ref{physicalsection}.

The requirement of gauge slice invariance of the partition function
with no insertions can be restated by taking $\tilde V = 1$ in the above 
argument to get
\begin{align}
   0 = \int d^2 x\, \sqrt {\bar g}\,\expect {\delta_B b_{ij} \cdots}\delta\hat g^{ij}.
\label{vone}
\end{align}  
If this is violated, the theory has a conformal anomaly.  

Such an anomaly can also be restated a 
non-invariance of the effective action under a quantum 
version of the BRST transformation to be defined below.
In the Lagrangian formalism considered here, this will be 
the most natural statement of the anomaly, usually encoded in 
the operator formalism in terms of $Q^2 \ne 0$.
Note that our transformation 
$\delta_B$ acts on 
the classical fields involved in the path integral and satisfies
$$
  \delta^2_B = 0
$$ 
by construction.    
Given this property of $\delta_B$, then in accordance with the
discussion of this issue in \cite{Witten}, 
any possible anomaly will be reflected in a failure of invariance
of the effective action $\Gamma$ under the quantum version of the 
BRST transformation $\delta_B$ which, following the
analysis of \cite{Weinberg}, is
expressed as 
$$
  \int d^2 x\, \sqrt {\bar g} \expect{\delta_B \Phi_i}_J {\partial\Gamma(\phi)\over \partial\phi_i} = 0,
$$
where $\Phi_i$ ranges over all the fields in the action, 
and $J$
denotes a set of sources for the fields $\Phi_i$.  The values of
the sources $J$
are fixed by the requirement that $\phi_i \equiv \expect{\Phi_i}_J$. 

It is now convenient to use the ghost-covariantized form (\ref{covBRST})
of the action, so that we can use manipulations similar to those
performed in section \ref{physicalsection}.  
The only possibly non-vanishing contribution to the above integral is then
given by 
the term
\begin{align}
  \int d^2 x \, \sqrt{\bar g}\expect{\delta_B\, b_{ij}} {\partial\Gamma\over \partial b_{ij}}
  &= 
    \int d^2 x \, \sqrt{\bar g}\expect{B_{ij}} {\partial\Gamma\over \partial b_{ij}} \nonumber\\
&= - {1\over 4\pi}\int d^2 x  \expect{\sqrt { g}\,T_{ij}(X, \chi, b, c, \rho,
     v, \omega)} {\partial\Gamma\over \partial b_{ij}} \nonumber\\
  &= - {1\over 4\pi}\int d^2 x \, \sqrt {\hat g}\,\expect{T_{ij}}_{\hat g} {\partial\Gamma\over \partial b_{ij}} \\
&=  {c\over 48\pi}\int d^2 x \, \sqrt {\hat g}\expect{T^i_{\phantom{i}i}}_{\hat g}\, {\partial\Gamma\over \partial b_\omega} + \cdots \nonumber\\
  &=  {c\over 48\pi}\int d^2 x \, \sqrt {\hat g}\,R_{\hat g}\, {\partial\Gamma\over \partial b_\omega} + \cdots,  \label{brstcondition}
\end{align}
where $b_\omega \equiv b_i^{\phantom{i}i}$, where $R_{\hat g}$ denotes the curvature, and where $c$ measures the combined anomaly
of the matter and ghost energy-momentum tensor
of the action $S$.  Since we are using the action (\ref{covBRST}), there 
is no $\beta$-$\gamma$ contribution.

Although this always vanishes for a flat world sheet, remember that
the effective action is a functional of all the fields in the original
action, including all configurations of $g_{ij}$.  It is not sufficient
for its variation to vanish only on a set of critical points in 
the space of $g_{ij}$.  Rather, the 
variation of the effective action must vanish on its entire domain, 
which then implies that $c$ must be zero in a consistent 
gauge-invariant quantization.  

We may simply relate the above BRST invariance condition to the 
previous condition of gauge slice invariance.  
Identifying 
$$
{\partial\Gamma\over \partial b_{ij}} \to \delta {\hat g}^{ij},
$$
in other words, calling the external source for $b_{ij}$, 
needed to define the effective action, by the name
$\delta {\hat g}^{ij}$, we see that this BRST-invariance 
condition is equivalent to the condition (\ref{vone}) for 
gauge slice invariance of the path integral.

We may also relate the above invariance condition to the 
anti-bracket formalism 
\cite{Weinberg} as follows.  Simply notice that, in our
gauge-fixed action, $\hat g^{ij}$ is
the external field coupling to the variation 
$\delta_B b_{ij} = B_{ij}$.
Considering the effective action $\tilde \Gamma$ as a functional 
also of $\hat g_{ij}$,
the above equation (\ref{brstcondition}) becomes
$$
  (\tilde \Gamma, \tilde\Gamma)
   \equiv
  \int d^2 x \,{\partial \tilde\Gamma \over \partial \hat{g}^{ij}}\,
     {\partial \tilde\Gamma \over \partial b_{ij}}
    =  {c\over 48\pi}\int d^2 x \, \sqrt {\hat g}\,R_{\hat g}\, {\partial\tilde\Gamma\over \partial b_\omega}.
$$
The condition that this should vanish 
is called the Zinn-Justin equation.

\section{D=26}
\label{anomalysection}

In \cite{myself}, it was shown that each matter field $X$, together with
its Pauli-Villars partners $\chi_i$, contribute $c = 1$ to the anomaly.
In this section we show that, in the current formalism, the ghosts,
together with their Pauli-Villars partners, contribute $c=-26$ to the 
anomaly.  

As in \cite{myself}, the 
anomaly can be read off from the coefficient multiplying the 
contact term
$$
  \expect {T_{z\bar z} T_{w\bar w}},
$$
which we will now calculate for the ghosts and their partners.  
Before we do so, starting from the action (\ref{covBRST}),
 we change variables  
\begin{align*}
c^\omega &\to c^\omega - \half\, \nabla_i c^i, \\
\omega &\to \omega - \half\, \nabla_i v^i,
\end{align*}
which has trivial Jacobian in the path integral for the same reason
that
$$
  \left(dx + \lambda \,dy\right) \wedge dy = dx \wedge dy, 
$$ 
and we obtain a ghost action of the form
\begin{align*}
   S_{\mathrm gh} &= \int \sqrt g\, b_{ab} \left(-\nabla^a c^b - \nabla^b c^a  
  +   g^{ab}\, \nabla_i c^i\right) - 2 \int \sqrt g\, b_{a}^{\phantom{a}a}\, c^\omega \\
 &\qquad    + \mathit{PV}.
\end{align*}
This action will be simpler than the original in conformal coordinates.

First, we note that the factor
\begin{align*}
  \int [db_a^{\phantom a}]\, [dc^\omega]\, 
        [d\rho_a^{\phantom a}]\, [d\omega]\,
        \exp\left({2 \int \sqrt {\hat g}\, b_{a}^{\phantom{a}a}\, c^\omega
         + 2 \int \sqrt {\hat g}\, \rho_{a}^{\phantom{a}a}\, \omega}\right)
\end{align*}
in the path integral is independent of $\hat g$.  This is trivially
shown by differentiating with respect to $\sqrt {\hat g}$ and using
$$
  \expect{b_{a}^{\phantom{a}a}\, c^\omega} = 
    - \expect {\rho_{a}^{\phantom{a}a}\, \omega}.
$$
As a result, these terms do not contribute to the anomaly and 
may be ignored.  
The remaining terms may be expressed in holomorphic coordinates
as 
\begin{align*}
  S = \half \int \biggl\{
           &- b_{zz}\, (2\bar\partial)\, c^z - b_{\bar z \bar z}\, (2\partial)\,
               c^{\bar z} \\
       &- c^z\, (2\bar\partial)\, b_{zz} - c^{\bar z}\, (2\partial)\,
               b_{z\bar z} \\
   &+ e^{-2\omega} \, \tilde m \, (b_{zz} b_{\bar z \bar z} - b_{\bar z\bar z} b_{zz})
- e^{4\omega}\,m \, (c^z c^{\bar z} - c^{\bar z}c^z)
  \biggr\}  + PV,
\end{align*}
where we have given small mass terms to the ghosts, to be taken to zero 
in the end, of the same form of the corresponding Pauli-Villars mass terms
in (\ref{covBRST}).  All dependence on $b_{z\bar z}$ cancels.   For now we 
have kept the Weyl factor $\omega$ defined by 
$g_{ij} = e^{2\omega}\, \delta_{ij}$ nonzero, so that from 
$$
  \delta_{\omega} S \equiv {1\over 4\pi} \int \sqrt g\, (2\delta\omega)\, T^i_{\phantom{i} i},
$$
we may read off 
$$
  T_{z\bar z} = 2\pi \,\left\{ -\half \, \tilde m\, b \bar b + m\, c \bar c\right\}
$$
at $\omega = 0$.  Here we abbreviated $b \equiv b_{zz}$, $\bar b \equiv b_{\bar z \bar z}$.

We may read off the propagator of $(b, \bar b, c, \bar c)$ from the action.
At the point of interest $\omega = 0$, it is given by
$$
  {1 \over 4\,\partial \bar\partial + \tilde m m}\left(
  \begin{array}{cccc}
     0 & m & 2\partial & 0 \\
     -m & 0 & 0 & 2\bar\partial \\
     2\partial & 0 & 0 & - \tilde m \\
    0 & 2\bar \partial & \tilde m & 0
  \end{array}
\right)
$$ 
We would like to calculate the ghost contribution
\begin{align}
   \expect{T_{z\bar z} T_{w\bar w}}
       = (2\pi)^2 \biggl\{
           &{1\over 4} \, \tilde m^2 \expect {b(z) \,\bar b(w)}
                                    \expect {\bar b(z) \, b(w)} \nonumber\\
           &+ m\tilde m \expect {b(z)\, c(w)} \expect {\bar b (z) \, \bar c (w)}  \nonumber \\
  & + m^2 \expect {c(z) \, \bar c(w)} \expect {\bar c (z) \, c (w)} \nonumber\\
    & + PV
        \biggr\}  \label {ghostcontact}
 \end{align}
The first term is 
\begin{align*}
  (2\pi)^2\, \left({\tilde m^2\over 4}\right)\, m^2
       \int {d^2 p\over (2\pi)^2}\, e^{-ip\cdot x}
     \int {d^2k\over (2\pi)^2}\, {1\over [k^2 + \tilde m m]}\,
     {1\over [(p - k)^2 + \tilde m m]} 
    + PV,
\end{align*}
which was calculated in \cite{myself} to give 
\begin{align}
    {2\pi\over 12}\, \partial\bar\partial\,\delta^2(z-w) \label{first}
\end{align}
in the limit $m, \tilde m \to 0$ and $M_\rho, \tilde M_v \to \infty$,
given that appropriate Pauli-Villars relations on the masses are satisfied.  
Similarly, the third term is 
\begin{align*}
  (2\pi)^2\,  m^2\, \tilde m^2
       \int {d^2 p\over (2\pi)^2}\, e^{-ip\cdot x}
     \int {d^2k\over (2\pi)^2}\, {1\over [k^2 + \tilde m m]}\,
     {1\over [(p - k)^2 + \tilde m m]} 
    + PV,
\end{align*}
which gives
\begin{align}
    {8\pi\over 12}\, \partial\bar\partial\,\delta^2(z-w) \label{third}
\end{align}
in the limit.  
The second term can be written, with the notation $k \equiv k_1 + i k_2$ 
in numerators, as
\begin{align*}
  &(2\pi)^2\,  m\tilde m
       \int {d^2 p\over (2\pi)^2}\, e^{-ip\cdot x}
     \int {d^2k\over (2\pi)^2}\, {\bar k\over [k^2 + \tilde m m]}\,
     {p - k\over [(p - k)^2 + \tilde m m]} 
    + PV \\
 &\quad =
   (2\pi)^2\,  m\tilde m\, \half
       \int {d^2 p\over (2\pi)^2}\, e^{-ip\cdot x}
     \int {d^2k\over (2\pi)^2}\, 
     {\bar k (p - k) + k (\bar p - \bar k)\over [k^2 + \tilde m m]\,[(p - k)^2 + \tilde m m]} 
    + PV \\
  &\quad = 
 (2\pi)^2\,  m\tilde m\, \half
       \int {d^2 p\over (2\pi)^2}\, e^{-ip\cdot x}
     \int {d^2k\over (2\pi)^2}\,\biggl\{ - {1\over (p-k)^2 + \tilde m m}
   - {1\over k^2 + \tilde m m}  \\
  &\qquad\qquad\qquad\qquad \qquad \qquad\qquad\qquad \qquad + 
     {p^2 + 2\,\tilde m m\over [k^2 + \tilde m m]\,[(p - k)^2 + \tilde m m]} 
    + PV 
  \biggr\}.
\end{align*}
The first two terms are proportional to 
$$
  \tilde m m \, \ln {\Lambda^2\over\tilde m m} + PV
$$
which vanishes by the usual conditions on the Pauli-Villars masses and
statistics.  The Fourier transform of the last term may be rewritten
as \cite{myself}
$$
  (2\pi)^2\, \half\, \left({\pi\over 3}\right)\,
    {1\over 2\,(2\pi)^2} \left\{\int_{2m}^\infty d\mu\,c(\mu, m)\, 
     {\mu^4\over p^2 + \mu^2}\, \left[ {p^2\over m^2}+ 2\right] + PV\right\},
$$
where the spectral function $c(\mu, m)$ is given in \cite{myself}.  
To save space on notation, we have set $\tilde m = m$.  The 
contribution proportional to the $+2$
in the square brackets was calculated in \cite{myself} to give, in 
conjunction with the Pauli-Villars contributions, the result
\begin{align}
  {\pi\over 6}\, p^2. \label{secondsecond}
\end{align}
The term proportional to $p^2/m^2$ may be rewritten as
\begin{align*}
  &(2\pi)^2\, \half\, \left({\pi\over 3}\right)\,
    {1\over 2\,(2\pi)^2} \, p^2\left\{\int_{2m}^\infty d\mu\,c(\mu, m)\, 
     {\mu^4\over p^2 + \mu^2}\,{1\over m^2} \left[\mu^2 - {\mu^2 p^2\over p^2 + \mu^2}\right] + PV\right\} \\
   &\quad = {1\over 4}\, \left({\pi\over 3}\right)\,
    \,p^2 \int_1^\infty {3\over 2}\, {d\eta\over \eta^4}\,
    {1\over \sqrt{\eta^2 - 1}}\, \left[4\eta^2 - {4p^2\eta^2\over p^2 + 2 m^2 \eta^2} + PV  \right].
\end{align*}
The first term is independent of $m$, and will vanish in conjunction with 
the Pauli-Villars contributions by the usual
condition $\sum c_i = 0$ on the statistics.    
The second term vanishes for the 
Pauli-Villars fields in the limit $M\to \infty$, and only the 
ghost contribution remains, which for $m\to 0$ is given by
\begin{align}
   {1\over 4}\, \left({\pi\over 3}\right)\,
    \,p^2 \, \left( {3\over 2}\right) \int_1^\infty
     {d\eta\over \eta^4}\,
    {1\over \sqrt{\eta^2 - 1}}\, (- 4\eta^2)
     &= {1\over 4}\, \left({\pi\over 3}\right)\,
    \,p^2 \, (-4)\, \left({3\over 2}\right)
   \nonumber \\
   &= - {\pi\over 2}\, p^2. \label {secondfirst}
\end{align}
Adding (\ref{secondsecond}) and (\ref{secondfirst}), and taking the Fourier
transform, we obtain the result
\begin{align}
   {16\pi\over 12}\, \partial\bar\partial\,\delta^2(z-w) 
  \label{second}
\end{align}
for the second term in (\ref{ghostcontact}).  Adding the three contributions
(\ref{first}), (\ref{third}) and (\ref{second}), we finally obtain, for the
ghosts and their partners
$$
    \expect{T_{z\bar z} T_{w\bar w}} = {26\pi\over 12}\, \partial\bar\partial\,\delta^2(z-w).
$$
From this, we can read off \cite{myself} the ghost contribution 
$$
   c = -26
$$ 
to the anomaly.

\section{Conclusion}

In this article we  discussed a 
covariant functional integral approach to the quantization
of the bosonic string.
We showed that interesting operators could be renormalized as 
true tensors, independently of whether the theory has a Weyl anomaly.
As a result, issues related to the anomaly could be isolated more clearly. 
This method of operator renormalization is 
in principle of wider applicability to 
covariant theories that are not Weyl invariant.  
  
Also of wider applicability in generally covariant theories
is our construction of a class of background-independent 
path integral measures, as well as the
construction of the BRST action from first 
principles and the discussion of its background invariance.
Interestingly, the BRST action could not be written in a 
background-independent way, although we showed that
one has some freedom in shifting
the dependence on a background metric from one set of terms 
to another.  However, it should be emphasized that this background
dependence is completely innocuous, since no dependence 
on the background metric remains in the physical results.  

Overall, the familiar string theory results are all 
reproduced in the current formalism.  What is interesting,
and instructive, 
is that they are encoded in the formalism somewhat differently
from the usual approaches.  The formalism of this paper 
separates the issue of operator renormalization very 
cleanly from the concern of anomaly analysis.

\section*{Acknowledgments}

    I would like to thank Dr. Miquel Dorca for very useful discussions.
    I would also like to thank Prof.\ Antal Jevicki and the Brown
    University Physics department for their support.


\begin{thebibliography}{99}
 \bibitem{myself}
   A. van Tonder,
   \textit{Coordinate-invariant path integral methods in conformal field theory}, to appear in \textit{International Journal of Modern Physics A}, 
  hep-th/0412031.
\bibitem{fujikawa1}
   K. Fujikawa,
   \textit{Path integral for gauge theories with fermions},
            \textit{Physical Review D} \textbf{21} No. 10 (1980), 2848-2858.
\bibitem{fujikawa2}
   K. Fujikawa,
   \textit{Path integral of relativistic strings},
            \textit{Physical Review D} \textbf{25} No. 10 (1982), 2584-2592.
\bibitem{PV}
   W. Pauli and F. Villars,
   \textit{On the invariant regularization in relativistic 
     quantum field theory},
   \textit{Reviews of Modern Physics} \textbf{21} (1949), 434-444.
 \bibitem{vilenkin}
   A. Vilenkin,
   \textit{Pauli-Villars regularization and trace anomalies},
            \textit{Il Nuovo Cimento} \textbf{44A} No. 3 (1978), 441-449.
\bibitem{anselmi}
  D. Anselmi,
   \textit{Covariant Pauli-Villars regularization
     of quantum gravity at the one loop order},
            \textit{Physical Review D} \textbf{48} (1993), 5751-5763.
\bibitem{nakahara}
   M. Nakahara,
   \textit{Geometry, Topology and Physics},
            Institute of Physics Publishing, London, 1990.
 \bibitem{eguchi}
   T. Eguchi,
   \textit{Conformal and current algebras on a general Riemann surface},
      in Proceedings of the First Asia Pacific Workshop on High Energy 
     Physics: \textit{Conformal Field Theory, Anomalies and Superstrings},
        B.E. Baaquie, editors \dots [et al.],
            World Scientific, Singapore (1987).
\bibitem{ooguri}
   T. Eguchi and H. Ooguri,
 \textit{Conformal and current algebras on a general Riemann surface},
  \textit{Nuclear Physics} \textbf{B282} (1987), 308-328.
\bibitem{BPZ}
 A.A. Belavin, A.M. Polyakov and A.B. Zamolodchikov,
 \textit{Infinite conformal symmetry in two-dimensional quantum field theory},
  \textit{Nuclear Physics} \textbf{B241} (1984), 333-380.
 \bibitem{difrancesco}
   P. Di Francesco, P. Mathieu and D. S\'en\'echal,
   \textit{Conformal Field Theory},
            Springer-Verlag, New York (1996).
 \bibitem{polchinski}
   J. Polchinski,
   \textit{String Theory},
       Vol. 1,
       Cambridge University Press, Cambridge (1998).
\bibitem{henkel}
   M. Henkel,
   \textit{Conformal Invariance and Critical Phenomena},
       Springer-Verlag, Berlin Heidelberg (1999).
\bibitem{alvarez}
   A. Alvarez-Gaum\'e, C. Gomez, G. Moore, C. Vafa,
 \textit{Strings in the operator formalism},
  \textit{Nuclear Physics} \textbf{B303} (1988), 455.
 \bibitem{gawedzki}
   K. Gaw\c{e}dzki,
   \textit{Lectures on conformal field theory},
      in \textit{Quantum Fields and Strings: A Course for Mathematicians},
        Vol. 2, P. Deligne, editors \dots [et al.],
            American Mathematical Society, Providence (1999).
\bibitem{cappelli1}
            A. Cappelli and J.I. Latorre,
            \textit{Perturbation theory of higher-spin conserved currents
                     off criticality},
            \textit{Nuclear Physics} \textbf{B340} (1990), 659-691.
\bibitem{cappelli2}
            A. Cappelli, D. Friedan and J.I. Latorre,
            \textit{c-Theorem and spectral representation},
            \textit{Nuclear Physics} \textbf{B352} (1991), 616-670.
\bibitem{forte}
            S. Forte and J. Latorre,
            \textit{A proof of the irreversibility of renormalization 
                    group flows in four dimensions},
            \textit{Nuclear Physics} \textbf{B535} (1998), 709-728.
\bibitem{FP}
  L.D. Faddeev and V.N. Popov, 
   \textit{Physics Letters} \textbf{B25} (1967), 29.
\bibitem{BRST}
  I.V. Tyutin, 
  \textit{Lebedev preprint FIAN} \textbf{39} (1975), unpublished;
   C. Becchi, A. Rouet and R. Stora, \textit{Ann. Phys} \textbf{98}
    (1976), 28.
\bibitem{Witten}
  E. Witten, \textit{Dynamics of quantum field theory},
   in \textit{Quantum Fields and Strings: A Course for Mathematicians},
    \textbf{Vol 2}, Pierre Deligne, editors \dots [et al.],
     American Mathematical Society, 1999.
\bibitem{Weinberg}
   S. Weinberg, 
    \textit{The quantum theory of fields},
    \textbf{Vol 2},
    Cambridge University Press, Cambridge, UK, 2000.
\bibitem{myselffuture}
  A. van Tonder, forthcoming.  
 
    \end{thebibliography}
\end{document}